\documentclass{aa}  

\usepackage{graphicx}
\usepackage{txfonts}
\usepackage{float}
\usepackage{natbib,twoopt}
\usepackage{subfig}
\usepackage[breaklinks=true]{hyperref} 
\bibpunct{(}{)}{;}{a}{}{,}             
\begin{document}

   \title{Distinguishing ram pressure from tidal interactions: the Size-Shape Difference (SSD) measure}


   \author{R. Smith
          \inst{1,2}
          \and
          S. Tonnesen\inst{3}
          \and
          K. Kraljic\inst{4}
          \and
          P. Calder\'on-Castillo\inst{1}
          \and
          A. Marasco\inst{5}
          \and
          Y. Jaffe\inst{6,2}
          \and
          B. Vulcani\inst{5}
          \and
          B. M. Poggianti\inst{5}
          }

   \institute{Departamento de Física,
Universidad Técnica Federico Santa María,
Vicuña Mackenna 3939, San Joaquín, Santiago de Chile\\
              \email{rorysmith274@gmail.com}
         \and
            Millenium Nucleus for Galaxies (MINGAL)
         \and
             Center for Computational Astrophysics, Flatiron Institute, 162 5th Ave, New York, NY 10010, USA\\
             \email{stonnesen@flatironinstitute.org }
         \and
            Observatoire Astronomique de Strasbourg, Universit\'e de Strasbourg, CNRS, UMR 7550, F-67000 Strasbourg, France
         \and
            INAF-Osservatorio Astronomico di Padova
vicolo dell'Osservatorio 5
35122 Padova, Italy
         \and
           Departamento de Fisica, Universidad Tecnica Federico Santa Maria, Avenida España 1680, Valparaíso, Chile \\
             }

   \date{Accepted to A\&A June 16, 2025}

 
  \abstract
   {In dense environments, disk galaxies can be subjected to tidal interactions with other galaxies and/or ram pressure stripping. Some morphological features are clearly associated with one or the other interaction (e.g. tidal bridges versus long one-sided linear gas tails). But, under certain circumstances, both mechanisms can result in morphological features that could be easily confused, such as lopsided or asymmetric disks and unwinding spiral arms.}
   {Our aim is to develop new measures for application to asymmetric galaxies of this type that distinguish gravitational-only tidal interactions from ram pressure stripping cases, and that can be applied directly to simulations, and potentially to observations.}
   {We define a new measure for galaxies called the Size-Shape Difference (SSD) measure. This measure is sensitive to differences in the size and shape of a younger stellar population ($<200$~Myr) compared to that of an intermediate age stellar population (200--400~Myr). We use numerical simulations of galaxies undergoing gravitational-only tidal interactions and/or undergoing ram pressure stripping to test the SSD measure. }
   {Because ram pressure tends to directly alter the gas distribution, the younger stellar population (which best traces out the gas distribution) tends to change shape and morphology with respect to the intermediate age population. The SSD measure is sensitive to this change, and we find it can effectively distinguish between ram pressure and gravitational-only tidal encounters. In fact, we find it is even more effective at identifying cases where a combination of a tidal interaction and ram pressure has occurred together, as may arise in dense environments. As tidal interactions tend to enhance the spiral structure in disk galaxies, the effectiveness of the SSD measure is further enhanced when combined with a measure of the strength of the spiral arms.}
   {}


   \maketitle
%

\section{Introduction}

When disk galaxies fall into dense environments, such as clusters and groups, they can find themselves confronted with a number of environmentally related physical mechanisms that act to transform them morphologically \citep{Dressler80}. 
The motion of the disk galaxy through the hot ionised gas that is trapped within the potential well of the dense environment can impose a hydrodynamical ram pressure on the leading edge of the gas disk of the galaxy, compressing it, reshaping its gas disk \citep{GunnGott1972}. As young blue stellar populations tend to be located within the disk gas, the fuel for their formation, this in turn can reshape the optical morphology of the galaxy. Some disk gas can be stripped away into tails that trail downwind from the galaxy disk in a process known as ram pressure stripping. In some cases, star formation can arise extragalactically within the stripped gas. The resulting morphology of such a galaxy has been described as a `jellyfish galaxy' where the body of the jellyfish is the galaxy disk, and the tentacles of the jellyfish are from the star formation in the streams of stripped gas (e.g., \citealt{owen06,Cortese2007,sun07,yoshida08,smith10,hester10,owers12,fumagalli11,ebeling14,mcpartland16,poggianti19}).  

However, the environmental effects are not limited to being hydrodynamical only. The deep potential well of the cluster as a whole can subject infalling disk galaxies to tidal forces. Furthermore, the cluster is filled with other satellite galaxies that may interact with the infalling disk galaxy in a process known as `harassment' \citep{moore98,gnedin03b,gnedin03,mastropietro2005,smith2010,smith2016}.

It is now well confirmed that star forming galaxies suffer reduced gas content \citep{davies73,huchtmeier76,haynes86,Solanes2001,Boselli2006} and quenching \citep{kennicut83,haanna,ha06,gavazzi13} in dense environments and, in general, ram pressure stripping seems to be a primary culprit (for recent reviews, see \citealt{Cortese2021} and \citealt{Boselli2022}). As a result, recently there has been a concerted effort to gather larger samples of jellyfish galaxies to study the process through which they pass. But, this is challenging due to the need for observationally expensive confirmation that they are undergoing ram pressure using radio observations (e.g.,\citealt{Roberts2021a,Roberts2021b}) or H-alpha/IFU (e.g., \citealt{Poggianti2017}). A number of optically selected jellyfish samples have recently become available (e.g., \citealt{poggianti16,Roberts2020,Vulcani2022,Salinas2024,Crossett2024}) but it can be difficult to fully confirm that they are jellyfish galaxies. Asymmetries in the dark matter halo are known to produce lopsidedness, asymmetries and tails that could easily be confused with ram pressure-like morphologies, as can tidal interactions with a nearby companion galaxy \citep{Zaritsky2013,Varela-Lavin2023}. 

An important example of this is so called `unwinding' of spiral arms by ram pressure, where the spiral arms become unwound on one side of a spiral galaxy and, in some cases, the opening angle of the spiral arms increases with radius. This morphology has been demonstrated to occur as a result of ram pressure stripping in \cite{bellhouse2021}. But it can also be seen in well known examples of tidal interactions, where a companion galaxy is clearly interacting (e.g. M51, \citealp{Dobbs2010}). 

{When collecting a sample of candidate jellyfish galaxies based on their morphologies, typically we first exclude galaxies showing clear and obvious merger features, such as tidal bridges connecting to a neighbour galaxy or shells, from the sample. Then, any remaining galaxies that present clear lopsidedness, asymmetries, and/or unwinding spiral arms could potentially be undergoing ram pressure. Or alternatively, they may simply have had a gravitational interaction with a neighbouring galaxy. Thus, ignoring this morphology risks excluding some real jellyfish galaxies, while including all examples of this type of morphology risks accidentally including galaxies undergoing tidal interactions instead of ram pressure. This can substantially change the fraction of candidate jellyfish galaxies in a sample. For example, in \cite{Vulcani2022}, the fraction of candidate jellyfish galaxies doubled when including unwinding morphologies alongside traditional stripping candidate morphologies within a population of blue, bright late-type galaxies within clusters. The problem is further compounded in dense environments, where crowding of the field makes it difficult to differentiate an interacting nearby companion from a by-chance projection effect. 
   
Furthermore, in dense environments it may also be undesirable to simply categorize galaxies into tidally interacting or ram pressure stripping. In fact, hydrodynamical cosmological simulations suggest that, frequently, these two environmental mechanisms arise in parallel \citep{marasco16}. Indeed, near the pericenter of a galaxy's orbit, both ram pressure and tidal effects might be expected to peak in strength. Therefore, there is additional value if a measure can be found that can detect the presence of ram pressure stripping, even it occurs in combination with a tidal interaction.

The aim of this study is closely related to this - we seek to develop new measures that are able to identify when ram pressure is involved in deciding the morphology of a disk galaxy that presents an asymmetric, lopsided and/or unwinding morphology (in the absence of clear merger features). Using controlled numerical simulations, we model two scenarios that can effectively generate these types of features. We study the case of pure ram pressure stripping, we compare it with the case of pure gravitational-only tidal interactions, and we even test combinations of the two mechanisms. We design and test out new measures on these simulations. While in this study we limit ourselves to designing and testing our new measure on these simulations, we do so with an aim to creating observational measures in the near future.

Our paper is structured as follows; in Section 2 we describe the set up of our galaxy models and simulations of the environmental mechanisms, in Section 3 we describe our new measures for detecting the presence of ram pressure, in Section 4 we show the results of applying these measures to the simulations. Finally, in Section 5 we summarize and draw conclusions.

\section{Set-up}
\subsection{Disk galaxy initial conditions}
\label{sec:DICE}
We build the initial conditions of our model disk galaxies using the publically available initial conditions set-up code {\sc{DICE}} \citep{Perret2014}. DICE builds a composite model galaxy including a dark matter halo, stellar disk, and gas disk component. By measuring the shape of the gravitational potential from the total model computed on a multi-level grid, it assigns the positions and velocities of dark matter and star particles to try to produce an overall model where the different components are in equilibrium with each other. We find this produces initial conditions that are already very stable. However, we also evolve our model galaxies in isolation to confirm their stability, and to provide an isolated control model for our models of environmental effects. 

The dark matter halo of the model galaxy has an NFW profile \citep{nfw}, with a virial mass of 3.2$\times$10$^{11}$~M$_\odot$ and concentration=12, and consists of 5$\times$10$^6$ dark matter particles. The stellar and gas disk are both modelled with flat radially exponential disks, truncated at 4 scalelengths. The stellar disk has a mass of 1$\times$10$^{10}$~M$_\odot$ and an exponential scalelength of 2.2~kpc, and consists of 2.5$\times 10^5$ star particles. The disk has a typical gas-to-stellar mass fraction of 0.125 \citep{catinella18}, and the gas disk's exponential scalelength is 3.7~kpc, which is 1.68 times the stellar scalelength \citep{Cayatte1994}.

\subsection{Simulation code}
Our initial conditions are evolved with the adaptive mesh refinement code Ramses \citep{Teyssier2002}, using the option of the ‘acoustic’ Riemann solver. We chose a box size of 280~kpc which is sufficient to contain the model disk galaxy out to its virial radius. The boundary conditions are outflow like, to allow the flow of intracluster medium in wind tunnel runs to smoothly exit the box. The minimum refinement level is 6 and the maximum refinement level is 10, meaning a minimum cell size of 270 pc. Adaptive mesh refinement is controlled by multiple criteria, including the number of dark matter and star particles from the initial conditions that are found in a cell (we choose 50 for levels 7 to 9 and 30 for level 10), and the baryonic (stars and gas) mass in a cell which depends on the refinement level of the cell (at the highest level of refinement this correspond to a mass $\sim$4$\times$10$^3$~M$_\odot$). Additionally, refinement occurs to ensure the Jeans length is resolved to a factor of 4, up until the maximum refinement level. This choice of refinement criteria ensures that the gas and stellar disk is resolved to the maximum level of refinement out to the edge of the disk and also vertically out of the plane of the disk. Simulations are conducted for 745~Myr with a snapshot interval of 14.9~Myr. This is sufficient time for the secondary galaxy to complete a fly by and perturb the disk in tidal interaction runs, and for a gas tail to form and gas to be stripped down to the truncation radius in ram pressure stripping runs (see Fig. \ref{fig:fid_timeseq}). 

Our star formation and feedback recipe is based on recipes implemented in \cite{Perret2014}. Star formation occurs for gas above a density threshold of 0.1 H/cc, and follows a Schmidt relation dependency on gas density with a star formation efficiency of 0.03. This choice of parameters ensures our isolated model galaxy forms stars at a rate that is consistent with the star forming main sequence at z=0 according to \cite{Speagle2014}. Supernovae feedback from OB stars is treated as follows. The OB type stellar populations that reach an age of 10~Myr are transformed into supernovae, and release energy, mass and metals into the nearest gas cell. The energy release is 10$^{51}$ erg per 10~M$_\odot$ supernovae, and the release is treated thermally. 20\% of the mass of star particle that goes supernova is returned to the gas with a yield of 0.1. 

Gas cooling occurs radiatively \citep{Sutherland1993}, assuming a gas metallicity of 0.15 solar metallicities (where one solar metallicity is defined as 12 + log(O/H) = 8.69), down to a temperature floor. The temperature floor is dictated by the Truelove condition \citep{Truelove1997} to guard against artificial fragmentation of the gas. {We use the default background ultraviolet heating model of Ramses for redshift=0 \citep{Haardt1996}.

%
%

\subsection{Wind tunnel model of ram pressure}
\label{sec:windtunnel}
In order to model the impact of ram pressure stripping on our model galaxy, we use a so-called `wind-tunnel' approach. The model galaxy is placed at the centre of the simulation volume and a hot ionised gas wind is fed into the simulation volume from a specified region, with a chosen density and wind speed. This mimics the motion of the disk galaxy through an ambient gas, such as the intracluster medium of a cluster. However, in a wind-tunnel model, the simulation volume is in the frame of reference of the galaxy, and so the galaxy remains motionless in the volume, while the intracluster medium flows up against the galaxy. We note that in a wind-tunnel model the environment interacts with the galaxy in a purely hydrodynamical manner.

In this study, the region from which the wind is fed in is located at 250-270 kpc in the x direction (the end of the simulation volume is at $x=280$~kpc), and covers the full $y$-$z$ face of cube (0-280~kpc in the $y$ and $z$ direction). The wind is fed in along the $x$ direction, with a fixed gas density, velocity, and 1 million Kelvin temperature (i.e., a constant wind model), resulting in edge-on-disk stripping. However, our main conclusions remain valid for ram pressure winds that are more inclined to the disk plane (see Appendix \ref{sec:app_winddir}). In our Fiducial model of Ram Pressure (`RP-fid' model), the wind speed is 1000 km/s, and the wind density is 1$\times10^{-4}$ H/cc. This is roughly equivalent to the density of the intracluster medium that arises for galaxies falling into the Virgo cluster at a radius of $\sim$600~kpc from the cluster center \citep{Roediger2007}.  However, for comparison, we consider two other constant wind speed cases, a weaker and stronger ram pressure model (`RP-weak' and `RP-strong', see Table \ref{tab:RPtable} for a list of the ram pressure wind parameters). We also consider a single case of a time-varying ram pressure wind (see Appendix \ref{sec:app_RPSevol}).

\begin{table}
\caption{Ram Pressure (RP) Wind-tunnel Models}             
\label{tab:RPtable}      
\centering                          
\begin{tabular}{c c c c}        
\hline\hline                 
RP model & Wind density & Wind Speed \\
\hline                        
   RP-weak & 10$^{-4}$ H/cc & 500 km/s \\      
   RP-fid & 10$^{-4}$ H/cc & 1000 km/s \\
   RP-strong & 10$^{-4}$ H/cc   & 2000 km/s \\
\hline                                   
\end{tabular}
\end{table}

\subsection{Tidal interaction models}

To model the effects of tidal interactions, we primarily focus on the interaction of our primary galaxy with a loosely bound neighbour galaxy. As mentioned in the Introduction section, it is well known that close interactions with a neighbour galaxy can drive the formation of an asymmetric, unwinding-like morphologies in a disk galaxy that could be confused with unwinding by ram pressure stripping (e.g., M51, \citealt{Dobbs2010}). Of course, tidal interactions could also arise from the potential well of the cluster itself. However, since the local tidal field of the cluster varies on spatial scales that are much larger than the typical size of spiral galaxies, disturbances are expected to be symmetric \citep{merritt83,bird90,Boselli2006,gnedin03b}, and thus unlikely to be confused with the effects of ram pressure.
 
To model the effect of a close interaction with a neighbor galaxy, we insert a live model of a secondary galaxy into the simulation volume. The secondary galaxy consists of a spherical NFW dark matter halo and an exponential stellar component. We note that we deliberately do not include a gas component in the secondary galaxy. In this way, the interaction with the primary galaxy is purely gravitational and non-hydrodynamic, in stark contrast with the wind-tunnel models described in Sect. \ref{sec:windtunnel}. As we will see, this type of gravitational-only interaction is able to generate an asymmetric disk and unwinding of the spiral arms, features that can also arise due to the effects of ram pressure, and in the absence of clear merger features such as tidal bridges linking to a neigbouring galaxy or shells. To set up the secondary galaxy, we make a copy of the primary galaxy but then scale down its mass to the required mass ratio between the primary and secondary galaxy. In the fiducial model, this mass ratio is 1:4 (i.e. the secondary galaxy is a quarter of the mass of the primary). This scaling can be accomplished by randomly sampling one quarter of the particles of the primary galaxy and scaling down the velocities of the particles by the square root of the factor by which the mass was reduced. We confirm that this procedure results in a stable secondary galaxy by evolving it in equilibrium.

The primary galaxy is placed at the origin with no net velocity. We now insert the secondary galaxy into the simulation volume with the primary galaxy. In the Fiducial Tidal Interaction (TI-fid) model, the secondary galaxy is placed at ($x$,$y$,$z$)=(50,15,0)~kpc, with a velocity of ($v_x$,$v_y$,$v_z$)=(-150,0,0) km/s.  This results in an off-axis collision due to the initial $y$-position being non-zero, with a pericenter distance of 6.3~kpc. The direction of rotation of the disk is such that this tidal interaction occurs in the prograde direction, i.e., the motion of the neighbor past the primary is in the same sense as the direction of rotation of the primary's disk. The initial $v_x$ value of 150 km/s is also similar to the rotation speed of the disk, meaning that the disk rotation frequency matches well with the passing of the neighbor galaxy, resulting in a stronger resonance. This relative velocity is similar to those found in small groups of loosely bound galaxies. Given that a significant fraction of cluster members arrived at their cluster inside a host system (e.g., \citealt{McGee2009,delucia12,Han2018}), such tidal encounters could arise when galaxy groups infall into denser environments such as clusters.

\begin{figure*}
    \centering
    \includegraphics[width=180mm]{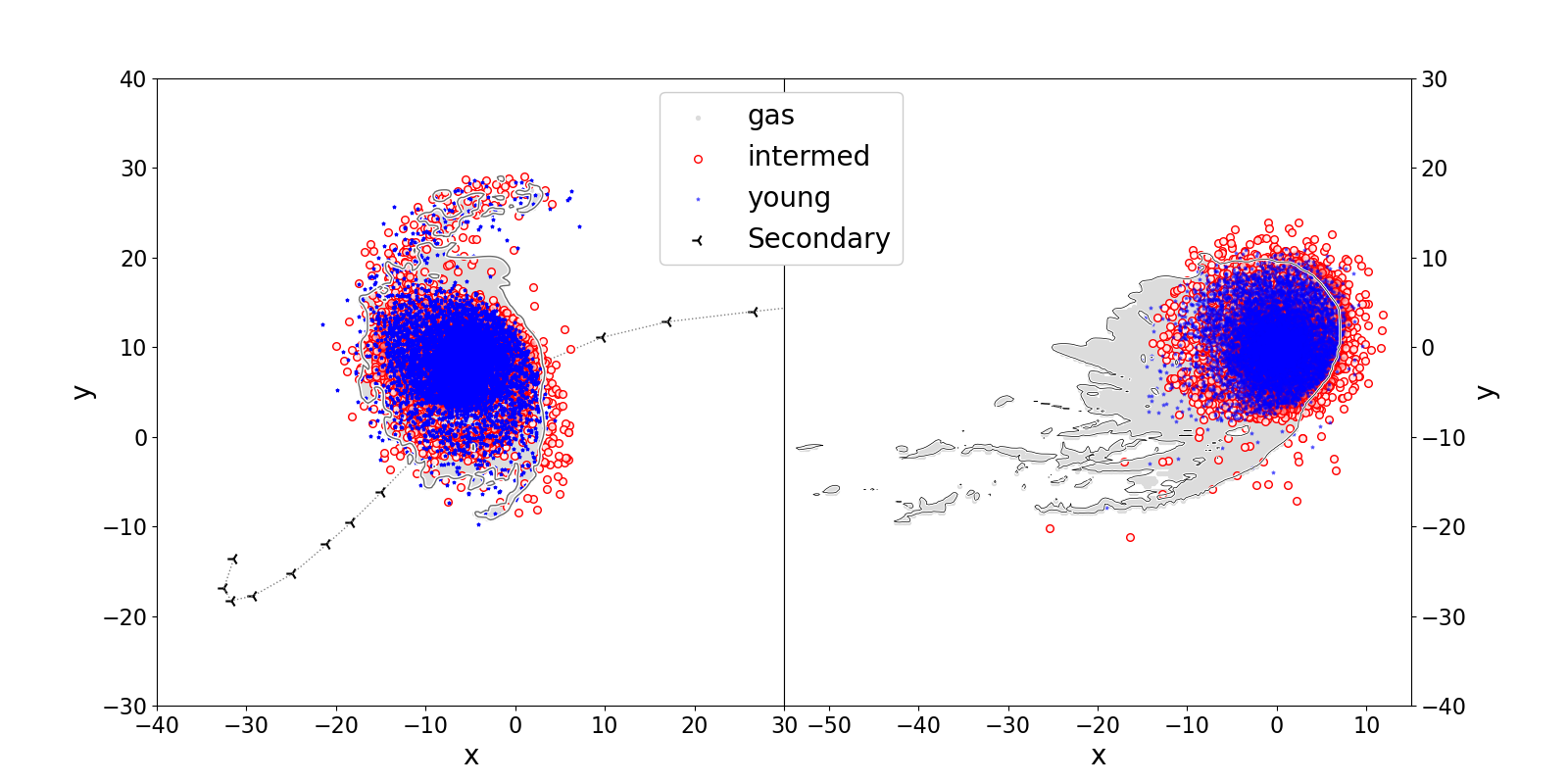}
    \caption{The x-y distribution of the young stars (blue points, age$<$200~Myr) and the intermediate age stars (red circles, age 200--400~Myr) at simulation time t$_\textrm{sim}$=626 Myr of the TI-fid model (left) and the RP-fid model (right). The x-y distribution of the disk gas (density$>0.01$~H/cc) is shown with gray shading and a darker gray outer contour line to highlight the extent of the disk gas. In the left panel, the trajectory of the secondary galaxy is marked with a symbol at intervals of 37 Myr. In the right panel, the ram pressure wind blows from right to left, along the x-axis.}
    \label{fig:agediff}
\end{figure*}

However, we also consider some other tidal interaction models that we find can mimic some of the features of a ram pressure morphology (e.g., asymmetry, one-sided spiral features, without a clear tidal bridge to a companion). This includes a model that is identical to `TI-fid', except the initial $y$-position is 25 kpc (instead of 15 kpc), resulting in a more distant tidal interaction (`TI-far') with a pericenter distance of 12.1~kpc (compared to 6.3~kpc in TI-fid). Additionally, we conduct another model that is identical to the `TI-fid' model, with the same pericenter distance, except the interaction occurs in retrograde instead of in prograde ($y$=-15 kpc instead of +15 kpc) labeled `TI-retro'). A summary of the main parameters of these tidal interaction models is given in Table \ref{tab:tidalmodels}.

We note that we also considered some more extreme tidal interaction models. These include a high speed interaction model (`TI-fast') where $v_x$ was increased to 800 km/s. In clusters, this is a typical interaction velocity between galaxies that were not associated with each other prior to infall into the cluster (i.e., a high speed fly-by interaction, \citealt{Moore1996}). However, this causes only a minor disturbance to the primary galaxy's disk, which tends to be very symmetrical, due to the short duration of the tidal interaction. We also conducted a much more major interaction where the secondary galaxy had equal mass to the primary galaxy (`TI-massive'). Neither of these models present morphologies that are likely to be confused with ram pressure. Nevertheless, for completeness, their results are presented in Appendix \ref{sec:app_extrememodels}. We also ran a simulation that is identical to the TI-fid model except we included a gas disk within the secondary galaxy as well. However, in this case the impact of the gas disks of the two galaxies generated a clear gas bridge filled with young stars after the first passage, linking the galaxies, and such a galaxy would be clearly identified as a galaxy merger. In such a situation, there would be less of a need to apply the SSD. We remind the reader that the aim of our study was to understand the origin of objects with asymmetries, lopsidedness and/or unwinding spirals but in the absence of obvious merger features, such as tidal bridges linking to a companion galaxy or shells.

Finally, we note that, while we model a close tidal interaction with a companion, we do not directly include a model for the tidal potential of the cluster. However, we numerically calculate that the tidal force at the edge of the galaxy disk dominates the cluster tidal force at all radii within the cluster, assuming a Virgo-like potential with an NFW halo of mass $5\times10^{14}$~M$_\odot$. At 600 kpc from the cluster center (where the intracluster gas density reaches the value assumed for the fiducial model), the tidal force of the galaxy dominates over the cluster potential out to a radius of 76 kpc from the galaxy center. This is much larger than the region in which we analyze the morphology of the primary galaxy's disk stars (typically $<20$~kpc). Therefore, in general, we do not expect there would be a significant change in our results by additionally including the cluster potential.

\section{Method}

\subsection{Key points for differentiating ram pressure from tides}
\label{sec:keypoints}
In Fig. \ref{fig:agediff}, we compare the Fiducial tidal interaction and ram pressure models, TI-fid on the left, RP-fid on the right. The models are compared at the same simulation time (t$_\textrm{sim}$=626 Myr). For a plot of the full time evolution of the TI-fid and RP-fid model, see Appendix \ref{sec:fid_timeseq}). 

The gray-scale shading in Fig. \ref{fig:agediff} indicates the gas distribution for all gas cells with density$>$0.01~H/cc (disk gas). We highlight the extent of the cold disk gas with a darker gray contour. In the RP-fid run, it is clear that ram pressure flowing from the right has stripped disk gas away from the leading edge into a tail that points downwind, towards the left of the figure. This gives the disk gas a comet-like appearance. Meanwhile, in the TI-fid run, the gas distribution does not replicate this one-sided comet shape, instead showing a prominent asymmetrical S-shape, matching that of the stars.

Young stars (with ages$<$200 Myr) are shown in blue, while intermediate age stars (200$<$age(Myr)$<$400) are shown in red. We will later consider alternative age bins for the young and intermediate age stars in Sect. \ref{sec:varyagebins}. The young stars tend to mimic the overall morphology of the gas, at radii where it is dense enough to meet the star formation threshold (0.1 H/cc). Hence, in the RP-fid run, the young stars show a tendency towards the comet-like shape (although less exaggerated), while in the TI-fid they show an S-shape morphology, somewhat similar to the disk gas.

Now, a key point can be seen which is at the heart of our new measurement for detecting the presence of ram pressure. In the TI-fid model, the intermediate age stars tend to share a similar overall morphology as the young stars. Meanwhile, in the RP-fid model, the morphology of the intermediate age stars differs from that of the young stars. The young stars are less extended than the intermediate age stars on the upwind side of the disk (towards the right of the panel), and more extended on the down wind side of the disk (towards the left of the panel).

One of the primary reasons for this difference that arises when ram pressure is present is that, in the several hundred Myr between the formation of the young and intermediate age stars, the gas distribution becomes more truncated on the leading edge (up wind). Thus, the young stars no longer form as far out in that direction as they did when the intermediate age stars were formed. Similarly, in the direction downwind, the gas disk becomes extended, enabling young stars to form further out than when the intermediate age stars were formed. Therefore, because ram pressure can directly affect the distribution of the star forming gas, the evolving gas distribution leads to different age gradients forming across the disk. 

Meanwhile, in the tidal models, the tidal disturbance generates a prominent S-Shape perturbation simultaneously in both the gas and stars. Because both components respond to the perturbation similarly, young stars tend to trace out a similar structure to the intermediate age stars, and the overall morphology of the two populations is more reminiscent of each other.

\begin{table}
\caption{Tidal Interaction (TI) Models}             
\label{tab:tidalmodels}      
\centering                          
\begin{tabular}{c c c c}        
\hline\hline                 
TI model & (x,y,z) & (v$_\textrm{x}$,v$_\textrm{y}$,v$_\textrm{z}$) & Mass ratio \\
  &  (kpc)  &  (km/s) & (Primary:Secondary) \\
\hline                        
   TI-fid & (50,15,0) & (-150,0,0) & 1:4 \\      
   TI-far & (50,25,0) & (-150,0,0) & 1:4  \\
   TI-retro & (50,-15,0) & (-150,0,0) & 1:4 \\
\hline                        
   TI-fast & (50,15,0) & (-800,0,0) & 1:4 \\
   TI-massive & (50,15,0) & (-150,0,0) & 1:2  \\
\hline                                   
\end{tabular}
\end{table}

We note that a secondary factor is also at play that causes the young stars to differ from intermediate age stars in the ram pressure models.  Once stars are formed, they will continue to orbit in the potential of the primary galaxy. For example, stars formed in the downwind gas tail of the galaxy, later return to the disk, maintaining the overall direction of rotation of the disk. However, this journey takes sufficient time that they may convert from being young stars to intermediate age stars by the time they have rotated around the disk. We see clear examples of this process in action in our ram pressure models (for example, see the lower-right panel of Fig. \ref{fig:fid_timeseq}). As a result, an over-density of intermediate age stars appear in the upper-right quadrant of the disk, causing the intermediate age disk to lose its round appearance and to appear to stretch upwards. Therefore, the shape of the intermediate age population can also be influenced by ram pressure, if sufficient time is provided for the perturbed initially young stars to evolve into intermediate age stars. And, this new shape of the intermediate age population continues to differ from that of the younger stars, but by following a different evolutionary route than we discussed previously.

These examples aim to highlight a key property at the heart of our new measure to detect the effects of ram pressure --  the ability for ram pressure to generate a differing morphology and size of the distribution of young stars compared to an intermediate age population. As we will see, the differences in morphology between the young and intermediate age populations are almost always larger in the presence of ram pressure than in our tidal interaction models. We aim to detect and sensitively quantify this difference using a new measure that we refer to as the `Size-Shape Difference' (SSD). 

In the following section, we describe the method to measure the SSD. But, before moving onto that description, we highlight one other useful difference that is visible in Fig. \ref{fig:agediff}. In our ram pressure models, we see a comet-like morphology may be induced in the newly formed stars, but we do not see evidence for strong enhancement of the spiral structure within the disk. The relatively smooth and flocculent spiral structure of the disk prior to ram pressure remains similarly smooth after ram pressure. However, in all our main tidal interaction models, the tidal perturbation induces spiral structure in both the stars and gas of the disk (i.e., the S-shape morphology we described before is an example of an induced grand-design spiral structure). Therefore, we note that a measure of the number of well defined spiral arms that are induced in the disk will likely provide additional information on if a strong tidal interaction has occurred. We will return to virtues of considering both the SSD and number of spiral arms in Sect. \ref{sec:2Dmeas}

\begin{figure*}
    \centering
  \subfloat{\includegraphics[width=155mm]{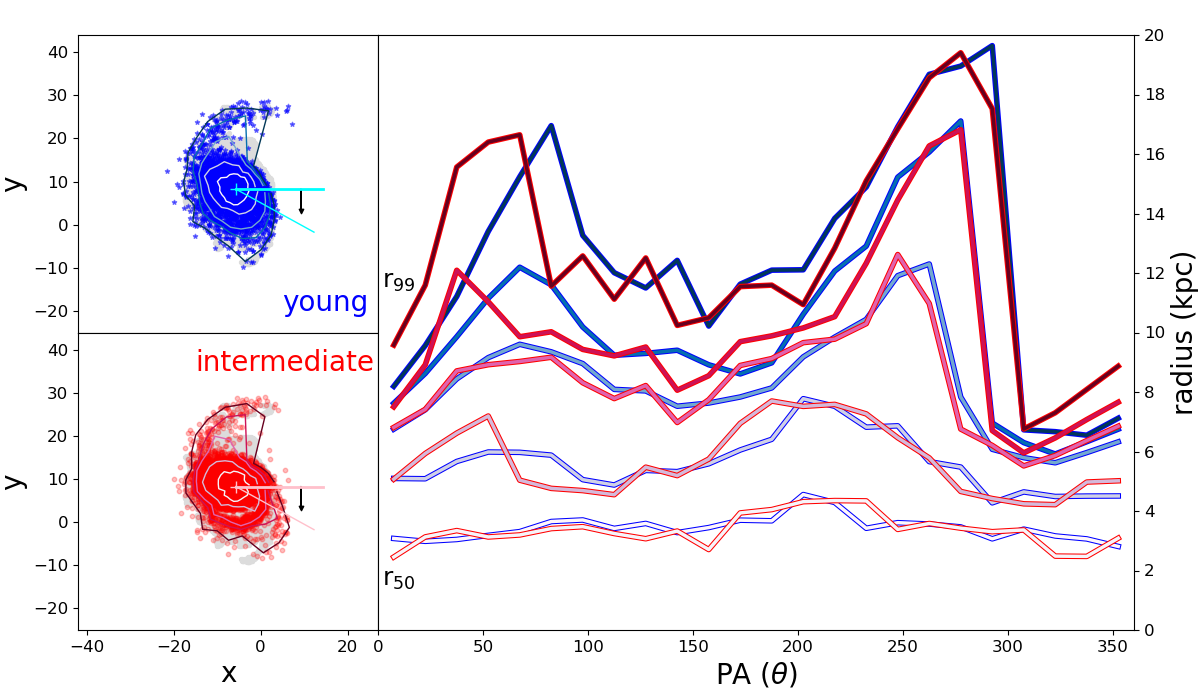}} \\
  \subfloat{\includegraphics[width=155mm]{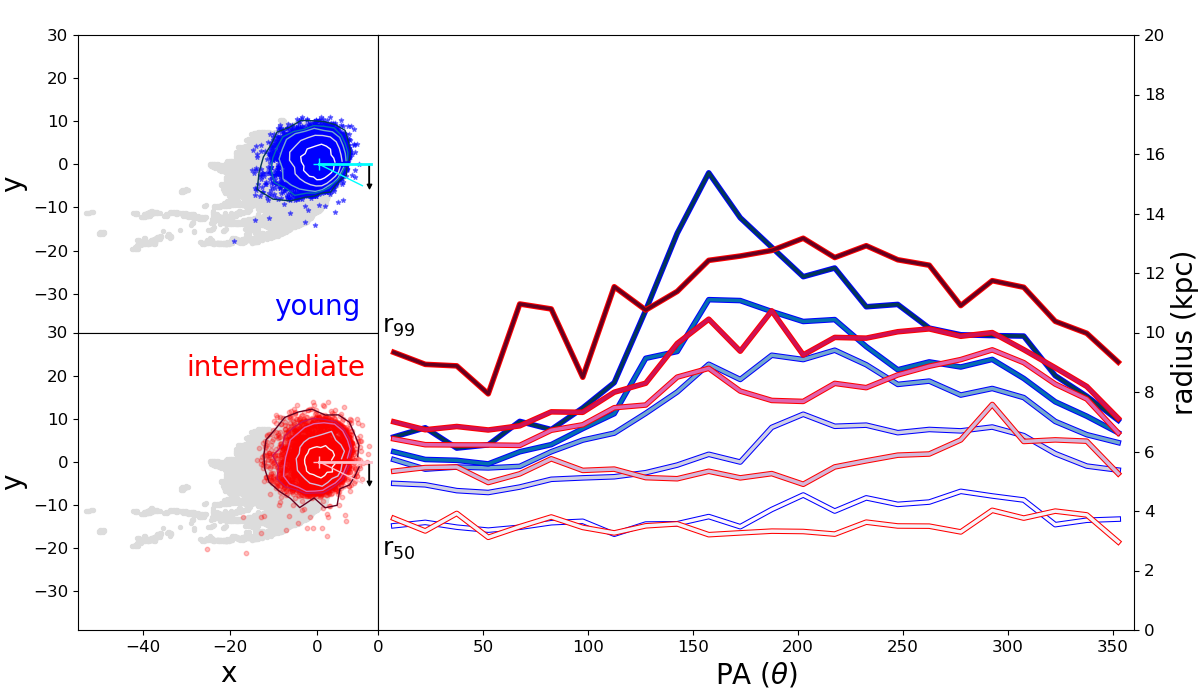}}
  \caption{Lagrangian radii containing 50, 75, 90, 95, and 99$\%$ (from bottom, r$_{50}$, to top, r$_{99}$) of the stars in 15 degree wide bins of position angle about the disk center (marked with a cross) at t$_\textrm{sim}$=626 Myr (to match Fig \ref{fig:agediff}). Upper panel: For the TI-fid model. An over-plot of the Lagrangian radii on an x-y plot of the young (upper-left sub-panel) and intermediate age (lower-left sub-panel) stars, with gray fill showing the extent of the disk gas, and right sub-panel: showing the Lagrangian radii as a function of position angle for the young (blue) and intermediate age stars (red). Position angles are measured anticlockwise (as indicated by the black arrow) from the horizontal line along the x-axis. Lower panel: same as for the upper panel except for the RP-fid model.} \label{fig:SSDmethod}
\end{figure*}

\subsection{How to measure the SSD value}
In this section, we describe how to measure the SSD at a single moment in time. However, the evolution of the SSD can also be measured while the simulation progresses, and we show examples of the resulting SSD evolution in Sect. \ref{sec:results}.

Here, we focus on the single moment in time shown in Fig. \ref{fig:agediff} (t$_\textrm{sim}$=626 Myr), starting with the TI-fid model. Our aim is to capture and quantify the differences between the young ($<200$~Myr) stellar population and intermediate age (200-400~Myr) stellar population. Starting with the young stars, we first trace out their shape and extent.

To begin with, we find the disk center. Using the intermediate age stars only, the location of the star with the maximum local density is used to mark the disk center. The fact that ram pressure can heavily perturb where young stars form drives our decision to use the intermediate age stars to find the disk center. This position for the disk center is now used for both the young and intermediate age stellar disks when calculating the SSD. 

Having found the disk center, we now divide up the disk of young stars into position angle slices about its center, in the plane of the disk (the $x$-$y$ plane). The range of angles in each slice must be sufficiently narrow to allow it to trace out the true shape of the disk but also not too narrow so that each bin would contain too few particles and become noisy. We chose angle slices that are 15 degrees wide, although we find our results are not highly sensitive to this choice.

\begin{figure*}
    \centering
    \includegraphics[width=140mm]{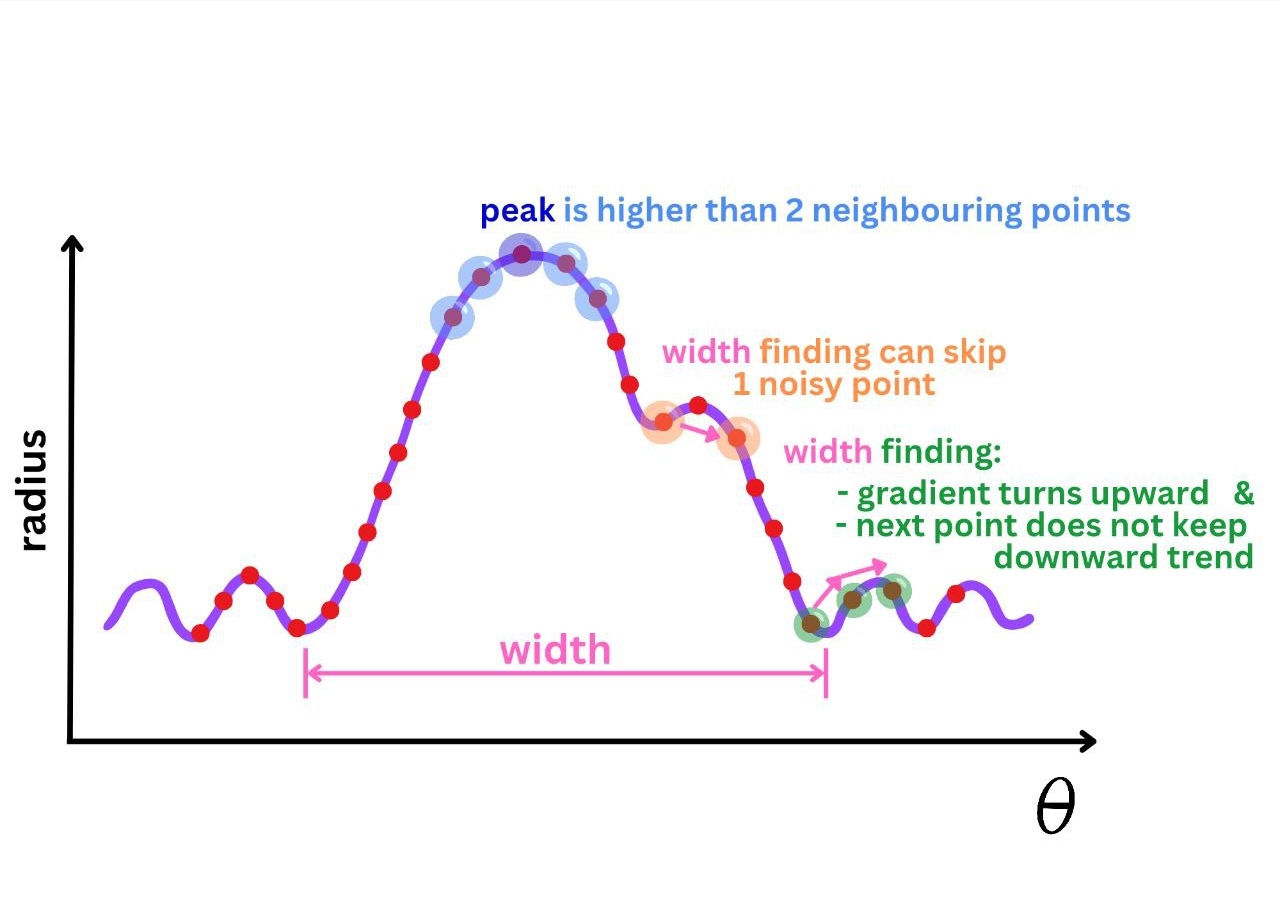}
    \caption{Cartoon schematic of the method used to identify statistically significant peaks in the 50, 75, 90, 95 and 99$\%$ Lagrangian radii plots (e.g. see right panel of Fig. \ref{fig:SSDmethod}). Peaks must be higher than two neighbouring points on either side, and must have a peak larger than a chosen signal-to-noise ($>$3.0). Additionally, the peak width must have a minimum angle width. See text in Sect. \ref{sec:Narmsmeasure} for further details.}
    \label{fig:Npeakcartoon}
\end{figure*}

Now, in each angle slice, we measure the radius containing different percentages of the total number of points in the angle bin -- so called `Lagrangian radii'. The percentages used are 50, 75, 90, 95 and 99$\%$. Thus the 50$\%$ Lagrangian radius encompasses half of the stars in that bin. But, if only the outermost edge of the disk is affected by the environmental mechanism, the 99$\%$ Lagrangian radius may be more sensitive to its impact. The exact choices of Lagrangian radii used are somewhat arbitrary, although we will demonstrate they are effective in Sect. \ref{sec:results}. However, we find there is value in combining multiple Lagrangian radii together. This is because, although the outermost Lagrangian radii are often the most sensitive to environmental effects, they are also frequently more noisy. Furthermore, even if only the outer disk is truncated by environmental effects, the innermost Lagrangian radii will shrink as well. This is because the Lagrangian radii are a tracer of a percentage of the \textit{total stars}, and if the distribution of the total stars changes, they are all affected to some extent. We test the impact of neglecting the two outermost Lagrangian radii in Appendix \ref{sec:no95}.

The Lagrangian radii for the TI-fid model are shown in the upper panel of Fig. \ref{fig:SSDmethod}. The upper-left sub-panel shows the Lagrangian radii overlaid on the x-y plane of the TI-fid galaxy's young stellar disk. The lower left-panel sub-panel is the same for intermediate age stellar disk of the TI-fid model. 

The right sub-panel shows the same Lagrangian radii but as a function of position angle measured around the disk's center (marked with a cross) in an anti-clockwise direction (as indicated by the black arrow) starting from the horizontal blue line along the x-axis. From bottom to top, the Lagrangian radii are the 50, 75, 90, 95 and 99$\%$ Lagrangian radii, respectively, with blue lines indicating the Lagrangian radii of the young stars, and red lines indicating the Lagrangian radii of the intermediate age stars. Prominent peaks at position angles of 50--100$^\circ$ and 250--300$^\circ$ highlight the locations of the induced spiral arms in both the young and intermediate age populations. The difference in the relative heights of the peaks emphasizes their asymmetry. It is notable that the position angle of the peak varies as we move from the inner- towards the outer-most Lagrangian radii, following the curve of the spiral arms.

Overall, the Lagrangian radii indicate that the shape and size of the young and intermediate age populations are quite similar. For example, the prominent peaks we noted previously are visible at similar position angles and with relatively similar size peaks as well. Some minor differences do arise but, as we will soon see, these are generally minor in comparison to the those of the ram pressure models. 

To quantify the differences at a single moment in time between the young and intermediate age populations, we calculate the SSD measure. This is the calculated as the modulus of the difference between each respective Lagrangian radius for the young and intermediate age stars, summed across all the position angles, and then summed into a single value that is a measure of the total of the differences in all the Lagrangian radii combined: 

\begin{equation}
SSD = \sum_{\theta=0^{\circ}}^{360^{\circ}} \sum_{i=1}^5 \big\lvert R(\theta)_{i}^{young}-R(\theta)_{i}^{intermed} \big\rvert
\label{SSDeqn}
\end{equation}
\noindent
where $R_1$, $R_2$, $R_3$, $R_4$ and $R_5$ is the 50, 75, 90, 95 and 99\% Lagrangian radii respectively, and the superscript shows whether it is the young or intermediate age stellar population. This single value is the measure of the SSD at that moment in time.

For the TI-fid model, we obtain an SSD value of 91.0 at the moment of time shown in Fig. \ref{fig:SSDmethod}. For comparison, the isolated model has an SSD value of 70.7 at the same moment in time. On its own, it is difficult to put these values in context. But, as we will see in Sect. \ref{sec:results}, the main tidal interaction models tend to give SSD values that are close or only slightly above those produced by the isolated control model, across the full span of time in which the tidal interaction drives a disturbance in the disk.

We now repeat this exercise for the RP-fid model as shown in the lower panel of Fig. \ref{fig:SSDmethod}. The truncation of the gas at the leading edge of the disk (the right hand side) results in the Lagrangian radii of the young stars being reduced with respect to their intermediate age counterparts in those directions, i.e., the blue curves are below the red curves for position angles near 0$^\circ$ or approaching 360$^\circ$. Meanwhile, on the trailing edge of the disk, young stars can be seen to appear at larger radii compared to the intermediate age stars, causing increased Lagrangian radii in those directions (for PA near 180$^\circ$). Given that we measure the absolute difference between the red and blue Lagrangian radii, the decreased Lagrangian radii in one direction and increased Lagrangian radii in the other direction work together to increase the SSD measure. For the RP-fid model, we receive an SSD value of 120.0, which is greater than the value for the TI-fid model and the isolated model at the same moment in time. In Sect. \ref{sec:results}, we will see that the SSD of RP models tends to be significantly higher than those of the isolated and tidal models at all times, throughout the period they interact with the environmental mechanism.

\subsection{How to measure the number of spiral arms}
\label{sec:Narmsmeasure}
As noted in Sect. \ref{sec:keypoints}, a common feature of the tidal interaction models is the formation of a well defined spiral structure, including the generation of a prominent double armed spiral. Although, depending on the parameters of the tidal interaction, one arm is frequently a different shape and/or length than the other. These show up as a double prominent peaks in the Lagrangian radii plots (50, 75, 90, 95, 99$\%$ radii) as can be seen in the upper panel of Fig. \ref{fig:SSDmethod} for the TI-fid model (at a position angle of 50--100$^\circ$ and 250-300$^\circ$ for both the young and intermediate age populations). Similarly, in the RP-fid model (lower panel of Fig. \ref{fig:SSDmethod}, the tail of young stars in the downwind direction produces a single peak in the radii plots in that direction (at a position angle of 130-200$^\circ$ for the blue outer-most Lagrangian radius).  Therefore, identifying the numbers of such environmentally induced peaks in the Lagrangian radii line plots of the young stellar population could provide additional information on the environmental mechanism at play, and complement the SSD measure. 

However, each Lagrangian radii line is not perfectly smooth, and subject to some noise. Therefore it is necessary to first filter out noisy peaks, so as to only count the number of genuine and significant peaks. To evaluate if a peak is significant or simply noise, we use a double criteria. We assume that a genuine peak must have a peak height with a minimum signal-to-noise (S/N$>3$), and it must have a minimum width (105$^\circ$) as the environmentally induced features tend to be quite broad in position angle, whereas noisy features are more narrow. In practice, we find this choice of values for the parameters is effective at helping to distinguish the isolated model from a model undergoing ram pressure or tides (as we will demonstrate in Sect. \ref{sec:2Dmeas}).

First peaks are identified in the Lagrangian radii versus position angle plots (e.g. right sub-panels of Fig. \ref{fig:SSDmethod}). We require that a true peak is higher than its neighbouring two points. The width of the peak is found by tracking down the peak on either side until the gradient turns upwards. Because of a noise, a single upward turn can arise by chance. Therefore, the gradient is allowed to turn upward, but if the next point continues the downward trend, then we will continue to track down the slope until the peak genuinely comes to an end, as shown in Fig. \ref{fig:Npeakcartoon}. Finally, to check if the peak is significant, we compare if its peak height passes the signal-to-noise criterion. The quantification of the noise is measured on the young stars of the isolated control disk simulation, evolved for the same length of time as the environmental models. The noise is given by the standard deviation of the values of the individual Lagrangian radii lines. We note that, even in the isolated model, the noise value evolves smoothly with time, steadily decreasing. However, we take this into account when computing the signal to noise of a peak in a model undergoing external environmental mechanisms by comparing them at the same time to the isolated model.

We visually check the success of this method for peak finding and find it is effective and stable over time, and for each of our models. We also experimented with a higher signal to noise cut, but found this could fail to detect some real environmentally induced peaks. Reducing the minimum peak width made the peak counting more noisy and less reliable (for example, finding more peaks than are actually present in the disks).

\begin{figure*}
    \centering
    \includegraphics[width=180mm]{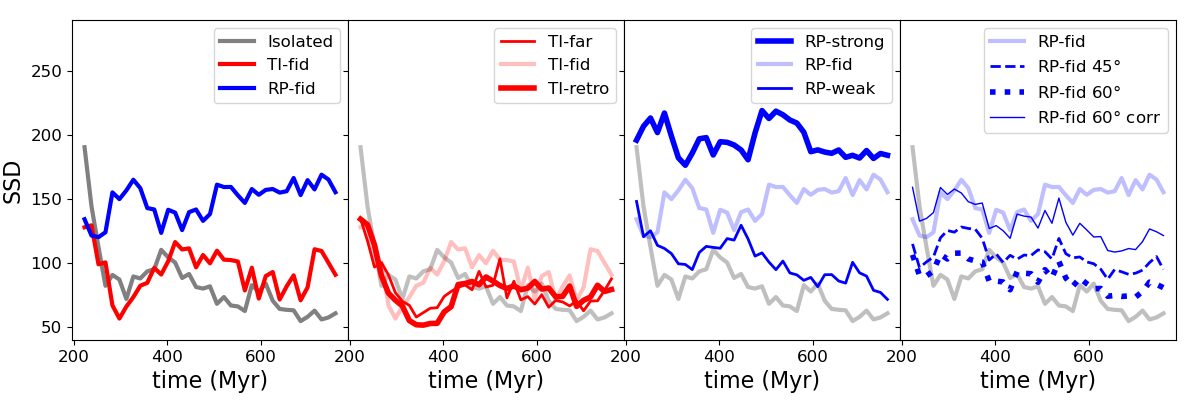}
    \caption{Comparison of the SSD evolution for various models. Far left panel: Comparison of the TI-fid (red) and RP-fid (blue) model. The isolated control model (gray) is shown for reference in all panels. Center left panel: Comparison between different tidal interaction models; the TI-fid (faint red), TI-far (thin red), and TI-retro (thick red) models. Center right panel: Comparison between models with different ram pressure strengths; the RP-weak (thin blue), the RP-fid (faint blue), and the RP-strong (thick blue) model. Far right panel: SSD evolution of the RP-fid model, observed with varying inclination angle to the spin vector of the disk. The face on disk view is RP-fid (faint blue), `RP-fid 45' has a 45$^\circ$ tilt away from face-on (blue dashed), and `RP-fid 60' has a 60$^\circ$ tilt away from face-on (blue dotted). The `RP-fid 60 corr' line (thin-blue line) applies the inclination correction described in the main text. In all cases, the ram pressure wind continues to be along the plane of the sky.}
    \label{fig:varystuff}
\end{figure*}

We note that our method for counting the number of peaks in the Lagrangian radii plots requires comparison with an identical control disk to evaluate the significance of the peaks, something which might not be available in an observational context or even in a fully cosmological simulation. In other cases, where such a control is not available, an alternative might be to quantify the number of spiral arms or to measure their opening angles (e.g., \citealt{bellhouse2021}), perhaps using a Fourier analysis (e.g., \citealt{Michea2021}) or other available software designed to quantify galaxy morphology (e.g. \textsc{galfit} \citealt{Peng2002}). Nevertheless, for our purposes, we find our approach is fast, easy to implement, and as we will show in Sect. \ref{sec:2Dmeas}, effective as an additional measure for differentiating cases of pure ram pressure from pure tidal stripping, or indeed identifying incidents of the two mechanisms combined.


\section{Results}
\label{sec:results}
\subsection{Evolution of the SSD for the RP-fid and TI-fid model}
\label{sec:SSDevol_fid}
In the left panel of Fig. \ref{fig:varystuff}, we plot the evolution of the SSD measure over the duration of the simulation. At early times, there are no intermediate age stars in the model as the simulation has not run for sufficient time for newly formed stars to enter the intermediate age bin. Hence, in this study, we wait until a few hundred Myr has passed before measuring the SSD (note, the x-axis starts at 200 Myr). Between t=200 to 220 Myr, the first stars formed in the simulation have sufficient age to begin appearing in the intermediate age bin. Initially, there are not very many in the intermediate age bin, which causes numerical differences to appear between the young and intermediate age populations, resulting in an artificially enhanced SSD value. But the SSD value rapidly falls, as the number of stars in the intermediate age bin increases, meaning the morphology of the intermediate age population is better represented.

For t$_\textrm{sim}>$220 Myr, there are sufficient numbers of intermediate age stars for the SSD measure to be reliable. The isolated model is shown in gray. The SSD value slightly declines with time, as the depletion of the gas disk causes the disk of young stars to slightly decrease in size over the duration of the simulation. The TI-fid model is shown in red, and the RP-fid model is shown in blue. It is notable that, after 250 Myr, the RP-fid model has a higher SSD value than both the TI-fid model and the isolated control disk at all moments in time. The factor by which the SSD value is higher for the RP-fid model varies from about 20\% to as much as 300\%. The mean increase of the SSD value between the RP-fid model and the isolated model is 194\%. Meanwhile the TI-fid model tends to stay closer to the value of the isolated control model. At later times, the SSD of the TI-fid model is still slightly enhanced over the isolated model as the young and intermediate age populations are not distributed identically, but the enhancement is much less than that seen for the RP-fid model. This is the first clear demonstration that the SSD is able to preferentially identify cases of ram pressure stripping. We note that the RP-fid model has a maximum peak of SSD value near t$\sim$300 Myr. This corresponds to the moment when the leading gas disk edge is most compressed, while the trailing edge displays a long tail of star forming gas (see second row of Fig. \ref{fig:fid_timeseq}).

We have also confirmed that our results do not qualitatively depend on the resolution and star formation threshold of our simulations. As described in Sect. \ref{sec:DICE}, our fiducial disk galaxy is refined to level 10, with a minimum cell size of 270 pc. At this resolution, we cannot resolve dense gas on smaller scales than 270 pc, therefore we chose a low value for the density threshold for star formation of 0.1 H/cc. To check that the strong enhancement in the SSD for the RP-fid model is not a strong function of these choices, we repeat the RP-fid model (and isolated control model) at level 12 (i.e., a minimum cell size of 68 pc), and with a density threshold for star formation that is increased to 1.0 H/cc. The SSD values of both the isolated and RP-fid model are found to be slightly smaller than in the higher resolution models (approximately 15-20\% lower than at our standard resolution). However, more importantly for this analysis, the SSD value of the RP-fid model is increased by a factor of approximately 3 compared to the isolated model, in alignment with our results at our standard resolution. Therefore, we conclude that the strong enhancement in the SSD value due to ram pressure is genuine, and not an artificial outcome of our choice of star formation parameters.

We also tested the effect of removing the two outermost Lagrangian radii (the 95\% and 99\% Lagrangian radii) from the SSD measurement of the RP-fid model. We find this can reduce the difference between the isolated and ram pressure stripped SSDs by 22\% on average in the case of the RP-fid model. But, there is still a significant enhancement of the SSD value when ram pressure occurs, compared to the isolated model. For full details, see Appendix \ref{sec:no95}.

\subsection{Varying tidal interaction parameters}
\label{sec:difftides}
In the center left panel of Fig.\ref{fig:varystuff}, we plot the SSD evolution for the different tidal interaction models; TI-fid (faint red), TI-far (thin red) and TI-retro (thick red). For reference, we once again plot the isolated control model (in gray). Comparing with the far-left panel, we note that all of the tidal interaction models are found below the RP-fid model. The RP-fid model typically has an SSD value which is approximately double that of the isolated and tidal interaction models. And, in general, all three tidal interactions (the TI-fid, TI-far and the TI-retro model) all show quite similar SSD values as the isolated control model. This once again confirms the ability of the SSD measure to preferentially detect the ram pressure of the RP-fid model over the tidal interactions we consider, in most cases giving values that are consistent with a model that is completely isolated for the tidal interaction models. We also consider two additional tidal interaction models, TI-fast (a high speed encounter) and TI-massive (a 1:2 mass ratio encounter) in Appendix \ref{sec:app_extrememodels}, from which we draw broadly consistent conclusions. Therefore, a weak response of the SSD to tidal interactions seems to arise in general, independent of the specific orbital parameters of the tidal interaction.

\subsection{Varying RP strength}
\label{sec:varyRP}
In the center-right panel of Fig. \ref{fig:varystuff}, we now consider the response of the SSD measure with time for different ram pressure strengths, as indicated in the legend (various blue lines). For reference, the isolated control model is also included (gray). The RP-strong model shows a significantly higher SSD than all the other models, at all times. This model forms a short but prominent tail downwind at early times, but converts into a fairly concentrated truncated disk thereafter. Thus, the high values of SSD at later times are a primarily a result of the fact that each Lagrangian radius (50, 75, 90, 95 and 99$\%$) is smaller for the young stars than for the intermediate age stars due to the truncation of the star forming disk. The RP-weak model provides SSD values that are in between those of the RP-fid model and the isolated control model, as might be expected given the ram pressures are weaker than the RP-fid model. We note that in the RP-weak model, the gas disk is not heavily stripped inside of the stellar disk, and as a result the SSD value is only weakly perturbed from the isolated control case. This demonstrates that the SSD will only detect the influence of ram pressure if it is sufficiently strong to affect star formation inside the disk. Mild ram pressures, where only more extended outer disk gas is stripped, will generate similar SSD values to that of the isolated control model.

\subsection{Varying disk inclination of RP-fid model to line of sight}
We have so far focused on measuring the SSD using a line-of-sight where the disk appears face-on, with the disk lying in the x-y plane while we look down the z-axis, and the ram pressure flowing along the x-axis in the negative direction (e.g., see right panel of Fig. \ref{fig:agediff}). Given that real galaxies may have a range of disk inclinations to our line-of-sight, here we recalculate the SSD varying the disk inclination angle to our line-of-sight. This is accomplished by rotating the RP-fid model about the x-axis. We consider a 45$^\circ$ and 60$^\circ$ rotation about the x-axis. 

\begin{figure}
    \centering
    \includegraphics[width=95mm]{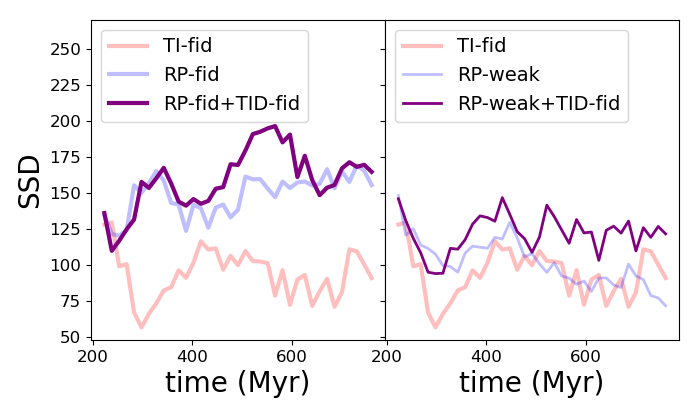}
    \caption{SSD evolution of the combined ram pressure and tidal interaction models (purple lines in both panels). Left panel: the TI-fid$+$RP-fid combination, and right panel: the TI-fid$+$RP-weak combination. For reference, the SSD evolution of the individual mechanisms (TP-fid, RP-fid, RP-weak) are included as faint lines (see legend for colors).}
    \label{fig:combinedRPTI}
\end{figure}

The resulting SSD measurement evolution is shown in the far right panel of Fig. \ref{fig:varystuff}. With increasing inclination, the deviation in the SSD from the face-on view of the isolated control model (in gray) is decreased. This is to be expected, as inclining the disk means that any differences in shape and size between the young and intermediate age stellar disks that occur along the y-direction are reduced by the cosine of the inclination angle. In other words, the SSD can only respond to the differences in shapes and sizes that are visible on the plane of the sky. However, given a known disk inclination angle of the disk to the line-of-sight, it is fairly straightforward to correct for inclination effects. In our case, we can simply divide the y-coordinates of the lagrangian radii by the cosine of the inclination angle to effectively cancel out the inclination effects. The resulting inclination-corrected SSD values can be seen in the right panel of Fig. \ref{fig:varystuff}, labelled `RPS-fid 60 corr'. From t$_\textrm{sim}$=200-550 Myr, it presents similar SSD values to the case of the face-on disk (labelled `RP-fid'). At later times the value is about 15$\%$ lower than the face-on value. Nevertheless, it remains significantly above the values of the tidal interaction models. This illustrates the effectiveness of this simple inclination correction that enables the SSD to maintain its value in detecting ram pressure stripping even in inclined disks.  

We also consider the case of a ram pressure wind that is inclined to our line-of-sight in the Appendix \ref{sec:app_winddir}. This is distinct from the previously described inclination tests, where the ram pressure wind was always edge-on, and it was the disk plane which was rotated to our line-of-sight. We find that, even in tests where the ram pressure wind direction is directly down our line-of-sight, the SSD is still effective in detecting the ram pressure, as long as the truncation of the gas disk is visible on the sky.

\subsection{Models of combined Ram Pressure and Tidal Interactions}
\label{sec:SSDevol_comb}
Given that we use controlled simulations, we are able to fully control the environmental mechanism the disk galaxy model is subjected to, and consider the case of pure ram pressure or pure tidal interaction. However, real galaxies that infall into dense environments can potentially face combinations of the two environmental mechanisms. Therefore, it is interesting to consider how the SSD measure responds under these circumstances. 

We simulate two different cases of combined ram pressure and tidal interaction. In both, the tidal interaction is set up as identical to that of the TI-fid model, but the ram pressure is either intermediate strength (like in RP-fid) or weak strength (like in RP-weak). For the parameters of these models, see Table \ref{tab:RPtable} and \ref{tab:tidalmodels}. Due to this choice of initial parameters, the ram pressure begins compressing the model galaxy's disk gas approximately 100~Myr before the pericentre passage of the secondary galaxy.

The left panel of Fig. \ref{fig:combinedRPTI} shows the SSD evolution of a combined model with intermediate strength ram pressure (purple line). For comparison, we include both the TI-fid and RP-fid curves (red and blue lines, respectively). It is interesting to note that the combined model shows a higher SSD than the other two models, even though they have equal strength environmental effects individually. In other words, when the mechanisms are combined, the SSD responds even more strongly than when the mechanisms acted separately.

\begin{figure}
    \centering
    \includegraphics[width=90mm]{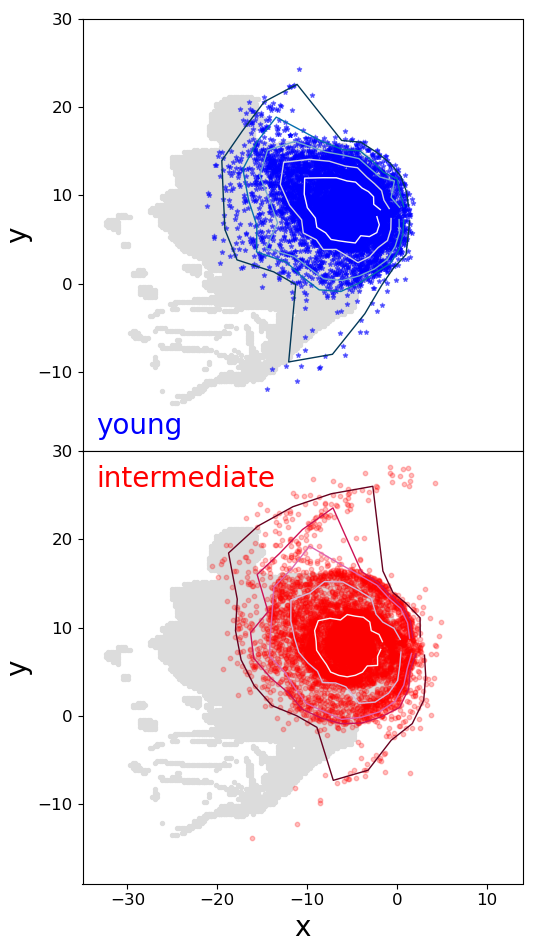}
    \caption{In the combined ram pressure and tidal interaction simulation (RP-fid $+$ TI-fid), the tidal interaction generates an S-shaped structure in the disk, and the ram pressure then peels back the gas from the extended spiral arms. The model is shown at time=611~Myr. For comparison the equivalent plots for the individual mechanisms can be seen at a similar simulation time in row 5 (left: TI-fid, RP-fid: right) of Fig. \ref{fig:fid_timeseq}. As a result, the young stars form outside the extended arms of the intermediate age stars, and the SSD is driven to even higher values than is seen in the separate RP-fid or TI-fid model (see SSD evolution in Fig. \ref{fig:combinedRPTI}; purple line of the left panel for the model shown here.)}
    \label{fig:combinedxyplot}
\end{figure}

A similar result is seen when we consider the combined model with a weak ram pressure (purple curve in right panel of Fig. \ref{fig:combinedRPTI}). Once again the combined model shows higher SSD values than for the TI-fid model or RP-weak model, meaning the SSD responds more strongly when the mechanisms are combined than when they act individually.

\begin{figure*}
    \centering
    \includegraphics[width=180mm]{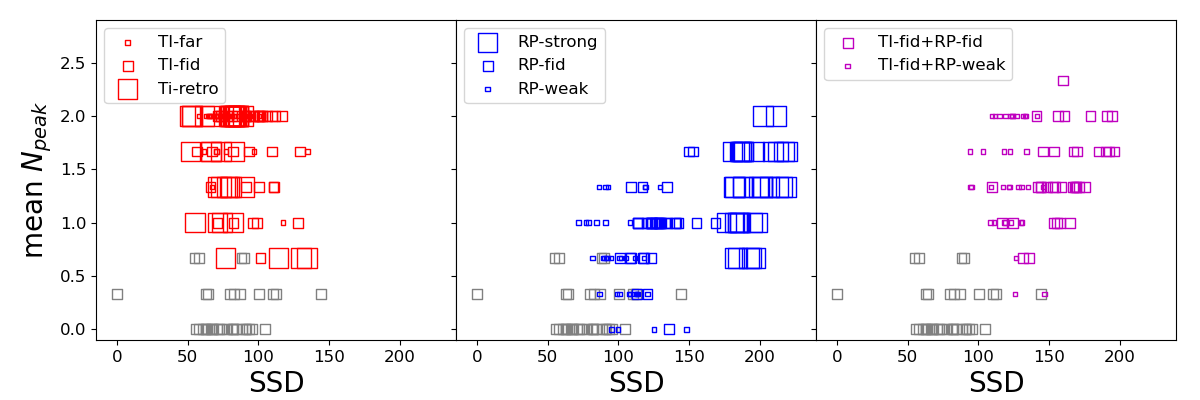}
    \caption{SSD on the x-axis versus mean number of peaks (N$_\textrm{peak}$) on the y-axis. A symbol is shown for each snapshot of the simulation with t$_\textrm{sim}>$220~Myr. Symbol size differentiates between the models (see legend in each panel). Left panel: the tidal interaction models TI-fid, TI-far, TI-retro show higher N$_\textrm{peak}$ values compared with the isolated galaxy (gray points, shown in all panels for reference). Center panel: the ram pressure models RP-weak, RP-fid, RP-strong are shown. Right panel: Models where tidal interactions are combined with ram pressure are shown (TI-fid$+$RP-fid and TI-fid$+$RP-weak).}
    \label{fig:SSDvsNpeak}
\end{figure*}

This property of the SSD measure is valuable, as it means we are still able to detect the occurrence of ram-pressure, even if it occurs in combination with a tidal interaction. In fact, under these circumstances, the SSD becomes even more sensitive to the ram pressure.

The reason why it becomes even more sensitive is because, in the combined cases, the tidal interaction initially generates an S-shaped spiral structure in the disk, like in the TI-fid case. However, subsequently, the ram pressure presses on the gas in the S-shape feature and distorts it, peeling back the leading spiral arm (see Fig. \ref{fig:combinedxyplot}) As a result, the location of the young stars differs significantly from the location of the intermediate age stars, and the SSD value is pushed up. In other words the tidal interaction generates a lot of disk structure that when distorted by ram pressure, creates even larger SSD values.

\subsection{SSD measure combined with N$_\textrm{peak}$}
\label{sec:2Dmeas}
So far, we have already seen that the SSD is effective at identifying ram pressure. Previously, we also noted that tidal interactions tend to induce (often asymmetric) S-shaped spiral structure in the models of the disks. This structure shows up as peaks in the Lagrangian radii versus position angle plots (e.g. top panel of Fig.  \ref{fig:SSDmethod}) and we devised a method to detect and count the numbers of these peaks in Sect. \ref{sec:Narmsmeasure}. 

Here we attempt to combine the benefits of both the SSD measure and the counts of the peaks by combining them into a 2D diagram (SSD on the x-axis, and mean number of peaks (N$_\textrm{peak}$) on the y-axis) as shown in Fig. \ref{fig:SSDvsNpeak}. The mean N$_\textrm{peak}$ is the mean of the number of peaks counted for the 90, 95 and 99$\%$ Lagrangian radii separately. This is motivated by the fact that the spiral structure is more clearly visible in the outermost Lagrangian radii. Also, as with the SSD measure, the use of a single Lagrangian radius can result in more noisy measurements. However, the exact choice to combine these Lagrangian radii is somewhat arbitrary, although we will demonstrate in this section that it works well in practice.  We measure N$_\textrm{peak}$ on the young stellar population so as it is more sensitive to the effects of ram pressure. 

In the left panel of Fig. \ref{fig:SSDvsNpeak}, we show the three tidal interaction models; TI-fid, TI-far, TI-retro, differentiated by symbol size (see legend). A point is shown for each snapshot in the simulation with t$_\textrm{sim}>$220~Myr (after which there are sufficient intermediate age particles for the SSD to be reliably measured, see Sect. \ref{sec:SSDevol_fid}). We compare their locations in the diagram with the isolated control model (gray points).

As was noted in Sect. \ref{sec:difftides}, the SSD provides quite similar values to the isolated control galaxy for all three of our tidal interaction models. We can see this in the left panel of Fig. \ref{fig:SSDvsNpeak} where the tidal interaction models do not deviate significantly from the isolated control model along the x-axis. But notably, the tidal models do offset from the isolated control model in the vertical direction, towards higher mean N$_\textrm{peak}$ values. They often reach a value of N$_\textrm{peak}$=2 whereas the control model typically has N$_\textrm{peak}$=0. This is because the tidal interactions induce an S-Shape spiral structure in a disk that would otherwise be quite smooth, with one peak for each arm of the S-shape.

In the center panel of Fig. \ref{fig:SSDvsNpeak}, we plot each snapshot of the ram pressure models, RP-weak, RP-fid, RP-strong (see symbol size in legend of that panel). As seen previously, the SSD responds strongly to the presence of ram pressure and the ram pressure models tend to shift horizontally to the right of the diagram, and the horizontal shift tends to increase with increasing ram pressure strength. We also see a vertical shift from the isolated control points, with the points generally grouping around N$_\textrm{peak}$=1 with some scatter to higher and lower values. An N$_\textrm{peak}$ value of 1 is what might be expected for a one sided tail of  young stars. But occasionally, values of 2 can genuinely occur when stars that formed in a ram pressure tail later fall back onto the disk.  Values of less than 1 can also arise as N$_\textrm{peak}$ is a mean of multiple Lagrangian radii and the inner most Lagrangian radii may be less affected than the outer most, especially when the ram pressure is weaker (hence why RP-weak tends to have lower N$_\textrm{peak}$ values than RP-fid and RP-strong). More importantly, by combining the SSD with N$_\textrm{peak}$, we are better able to detect weak ram pressures than we can from the SSD alone.

\begin{figure*}[h]
    \centering
    \includegraphics[width=180mm]{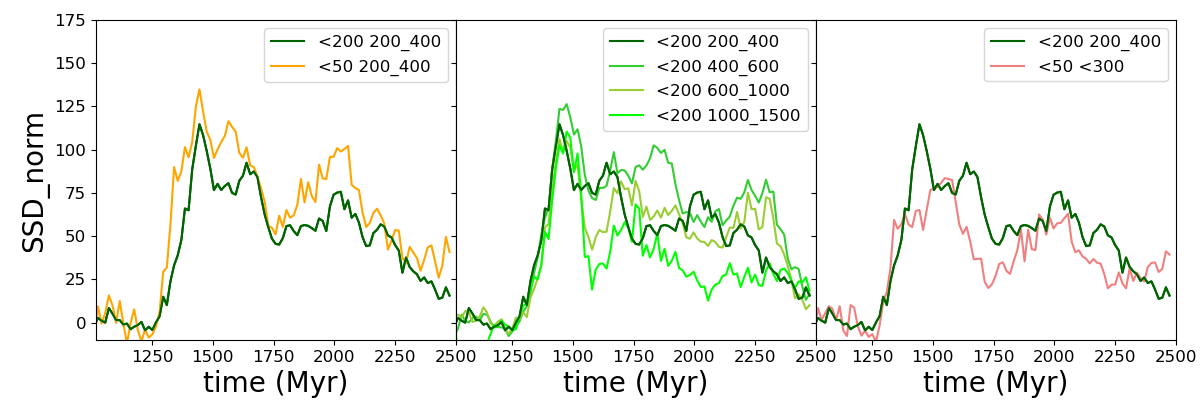}
    \caption{Normalised SSD evolution of a single model galaxy for differing choices of the age bins used to define the young and intermediate age populations. All curves are normalised so as the SSD$_{\rm{norm}}$ value is exactly zero at t$_{\rm{sim}}=1250$~Myr. The fiducial values are $<200$ Myr and 200--400 Myr respectively (dark green line in all panels). In all cases, the model galaxy is subjected to ram pressure (equivalent to RP-fid) that begins to affect the SSD value at t$_\textrm{sim}$=1250 Myr. Left panel: Young stars defined as $<50$Myr instead of $<200$Myr. Middle panel: Definition of the intermediate age bin is varied from 200-400 Myr up to 1000-1500 Myr (see legend in that panel). Right panel: The pink line is the SSD for a young population with age $<50$ Myr and an intermediate age population with age$<$300 Myr. These two age bins are no longer exclusive (i.e., young stars are found in the intermediate age bin).}
    \label{fig:SSDvariations}
\end{figure*}

Finally, in the right panel of Fig. \ref{fig:SSDvsNpeak}, we plot the location of the combined models; RP-fid$+$TI-fid and RP-weak$+$TI-fid (see symbol size in legend of that panel). Here we see the maximum benefit from combining the SSD and N$_\textrm{peak}$. As we saw in Sect. \ref{sec:SSDevol_comb}, the SSD responds strongly to the models with combined tidal interaction and ram pressure. Additionally, we see that the tidal interaction pushes the points upwards towards N$_\textrm{peak}$=2, cleanly separating them from the control models. Even when the ram pressure of the combined model is quite weak, the location of points in 2D space shows a greatly reduced overlap with the isolated control model compared to if we considered the SSD alone.

There is a large amount of support in the literature for the idea that strong tidal interactions can generate grand design spiral structure of the type that we would detect with N$_\textrm{peak}$=2 (e.g., \citealt[see also the review of \citealt{Dobbs2014}]{Byrd1992,Dobbs2010,Semczuk2017, Pettitt2018}). However, we note that there is still debate if all grand design spirals are tidally triggered or if some could arise without external perturbation (e.g. \citealt{Elmegreen1982}). Nevertheless, we emphasize that we are interested in understanding the origins of disk galaxies with clear asymmetries in their disks, features that are unlikely to spontaneously occur in isolated disks.

\subsection{The impact of stellar age on the SSD measure}
\label{sec:varyagebins}
We have so far focused on an SSD measure where young stars (assumed ages $<200$~Myr) are compared with intermediate age stars (assumed ages of 200 to 400 Myr). In this section, we consider what happens if we use alternative definitions of our young and intermediate age stars. 

Because we will sometimes choose an intermediate age bin that is much older, for these tests we first run our control model for 1~Gyr in isolation, in order to create a stellar disk with a broader range of ages. We then introduce the evolved model to a ram pressure wind with the same wind properties as the RP-fid model. The ram pressure begins perturbing the SSD values at t$_\textrm{sim}\sim$1250~Myr (see increase in SSD in panels of Fig. \ref{fig:SSDvariations}), and we continue the simulation until t$_\textrm{sim}$=2500 Myr. We also run an isolated disk over the same length of time as a control.

We note that, in the isolated control disk, over 2500 Myr of evolution, the SSD measure slowly and steadily evolves towards slightly higher values. This is because the isolated model slowly converts its gas disk into stars. As there is no hot halo or external gas feeding source to replenish the consumed disk gas, the star forming disk slightly shrinks over time. This causes the young stars to be smaller in extent than the intermediate age stars at later times, and the exact SSD value depends on our choice of age range definition for the young and intermediate age stars. This means that different age definitions can cause differing SSD values, due to the setup of the simulation rather than a reflection of the overall galaxy evolution, which makes comparing between the definitions more challenging.

To allow for a fairer comparison, we first measure the SSD value for the isolated disk evolved over the full 2500~Myr, for each age definition. We then normalise the SSD curves of the galaxy undergoing ram pressure by the value of the isolated control model with a matching age definition, at that time. This means that, just before the ram pressure wind begins to affect the star formation in the disk, the SSD values are normalised to be close to zero (e.g., at t$_\textrm{sim}\sim$1250~Myr in the panels of Fig. \ref{fig:SSDvariations}). The vertical offsets applied to the original SSD to ensure this occurs are listed in Tab. \ref{tab:ssdoffsets}. In this way, we can more clearly compare the size of the increase in the SSD due to the onset of ram pressure, without being confused by small differences in the SSD value even before the wind hits the disk.

\begin{table}
\caption{Table listing the vertical offsets applied to the original SSD values in order to ensure they have an SSD$_{\rm{norm}}$ value of zero at t$_{\rm{sim}}$=1250~Myr.}             
\label{tab:ssdoffsets}      
\centering                          
\begin{tabular}{c c c}        
\hline\hline                 
young & intermediate & SSD offset \\
\hline                        
$<$200~Myr & 200-400~Myr &  -48 \\
$<$200~Myr & 400-600~Myr &  -60 \\
$<$200~Myr & 600-1000~Myr &  -95 \\
$<$200~Myr & 1000-1500~Myr &  -126 \\
\hline
$<$50~Myr & 200-400~Myr &  -87 \\
$<$50~Myr & $<$300~Myr &  -67 \\
\hline                                   
\end{tabular}
\end{table}

\vspace{0.2cm}
\noindent
We consider several variations on the standard SSD measure:

\vspace{0.2cm}
\noindent
\textit{-Young stars defined as $<50$~Myr (instead of $<200$~Myr):}\newline This choice of young stars could be thought of as a rough proxy for H-alpha emission from young, massive recently formed stars. If so, in the case of applying the SSD measure to IFU observations, a H-alpha emission map could potentially be directly used (and which may be simpler to implement) to trace out the shape of the young stellar population, instead of trying to identify the young population from stellar ages (which adds in additional uncertainties), with the caveat that not all H-alpha emission may derive from young stars alone. 

The left panel of Fig. \ref{fig:SSDvariations} shows the SSD evolution for a young population with ages $<50$~Myr (orange) instead of the usual $<200$~Myr (dark green). We note that the SSD for the $<50$~Myr young disk is slightly more perturbed by the onset of ram pressure, likely because the $<50$~Myr disk better traces out the dense gas distribution at any moment in time. However, the differences are relatively small, meaning the SSD measure is not highly sensitive to the exact choice of age bin for the young stars. This could be of benefit, given that it is not trivial to separate different age stellar populations with a high degree of accuracy, even with IFU observations.

\vspace{0.2cm}
\noindent
\textit{-Intermediate age stars defined as 400-600 Myr, 600-1000 Myr or 1000-1500 Myr (instead of 200-400 Myr):}\newline
In the center panel of Fig. \ref{fig:SSDvariations}, we experiment with the limits of the age bin used for the intermediate age stars (see legend). Our fiducial choice is 200-400~Myr (the dark green line). After the onset of ram pressure, we see that all of the curves rise up to a similar maximum which occurs at about the same moment (t$_\textrm{sim}\sim$1450 Myr). This peak is the result of the compression of gas on the leading edge and formation of the downwind tail. The fact that all of the curves show quite similar behaviour until this time shows that the ability for the SSD to detect ram pressure during this phase is not highly sensitive to the choice of age bin for the intermediate age stars. Given the challenges in cleanly separating stellar populations with different ages, it is beneficial that we are not strongly tied to a specific choice of age bin in order to detect ram pressure with the SSD. 

Following the SSD peak, there is a reduction in the SSD. This is because the downwind tail of young stars rotates around and falls back onto the disk. Thus, the cometary appearance of the young stars gives way to a more round but highly truncated star forming disk. However, the reduced size of the star forming disk compared to the intermediate age stars means the SSD value remains elevated above the values of the isolated disk. This means the SSD measure remains able to detect a signature of ram pressure, even after the early cometary like morphology is no longer visible. 

It is unclear which choice of age bin is best for long term detection of ram pressure. The curves are quite noisy, as young stars formed in the tail fall back into the disk and may reemerge, both in the young stellar bin and later in the intermediate age bin. And, these stars will take longer to appear in the intermediate age bin depending on the exact definition of the bin. As a result, there is not a clear winner in terms of long term sensitivity, with the age bin with the strongest signal interchanging between the categories we consider as time evolves. However, the 1000-1500~Myr does consistently tend to give lower SSD values than the other categories during this post-cometary phase. This is because the young tail stars that fall back onto the disk maybe more settled by the time they appear in the 1000-1500~Myr age bin. Meanwhile, in the other age bins, the fall back stars appear as splashback-like features in images of x-y plane of the intermediate age stars. An example can be seen in the extended distribution of red points shown in the upper-right quadrant of the disk in the bottom-right panel of Fig. \ref{fig:fid_timeseq}). These features are found at very different position angles than the downwind tail of the young stars, and as a result the effect is to drive up the SSD value after the initial SSD peak (see SSD variations during the period t$_\textrm{sim}$=1600--2300 Myr in the panels of Fig. \ref{fig:SSDvariations}).

We note that, in the case of a heavily truncated gas disk, if we wait sufficient time since the disk truncation (i.e., beyond the upper age range of the intermediate stars), then both the young and intermediate age stars will exclusively be found inside the truncated gas disk. This could result in a similar size and shape for the two age bins, and thus a low SSD value at late times after ram pressure has occurred. An example of this can be seen in the case of face-on ram pressure stripping in Fig. \ref{fig:varywinddirection}, where the truncated gas disk is very symmetrical in the disk plane due to the perpendicular direction of the ram pressure wind. In this situation, in order to detect that ram pressure has occurred much earlier, it would be necessary to chose a definition of the intermediate age stars whose age range encompasses the time since the gas disk was truncated. This demonstrates that a comparison between the SSD values with differing definitions of the intermediate age population could potentially provide additional information on the time since changes in the gas disk truncation radius occurred.

\vspace{0.2cm}
\noindent
\textit{-Young stars $<50$~Myr and intermediate age stars $<300$~Myr:}\newline
In the right panel of Fig. \ref{fig:SSDvariations}, for the first time we consider the case where the intermediate age bin also includes the young stars as well (red line). This lack of exclusivity is intended to be a crude approximation for what might occur if a galaxy is observed in two filters, where the young stars are visible in both filters (for example, in the case of H-alpha and near-ultraviolet imaging). We plan to consider a more sophisticated approach to mock our simulations in a follow-up study. Nevertheless, this simple approach can give us some initial insights into what to expect. As we see in the right panel, the SSD using these new time bins still detects the onset of ram pressure. However, the SSD change is slightly reduced compared with the standard choice of $<200$Myr and 200-400 Myr for the young and intermediate age population (dark green line). This is to be expected as the difference in the young and intermediate age populations is inevitably reduced as they are made more similar by having some of the same stars in both bins. As a result, we can expect the SSD to be slightly more sensitive when the two stellar populations that are compared are mutually exclusive. This likely favors the application of the SSD to IFU observations, where attempts can be made to split the stellar population by age (although somewhat imperfectly). However, the difference between the dark green and red line is not substantial, which shows that the ram pressure can still be detected even when the age bins are not mutually exclusive. Therefore, in principle it is likely possible to use the SSD to detect ram pressure in broadband images of galaxies. We leave a detailed investigation of these issues to a dedicated follow-up study.


\section{Conclusions}
In certain environments, for example in the outskirts of clusters, galaxies may be subject to tidal interactions with neighboring galaxies or ram pressure stripping from their motion through a surrounding hot ambient gas. The morphological response of a galaxy's disk to a gravitational-only tidal interaction or ram pressure stripping can often appear similar. Both mechanisms are able to produce asymmetries, lopsidedness and/or unwinding spirals.  In the absence of clear merger features such as shells or tidal bridges linking to a companion galaxy, it may be difficult to determine which external mechanism has acted upon such galaxies. The aim of this study was to develop a new measure of whether ram pressure has occurred in galaxies with asymmetrical features where the origin would otherwise be unclear. As a combination of tidal interactions and ram pressure can often occur simultaneously in dense environments, ideally our new measure would be able to detect ram pressure even in cases where tidal interactions are also involved. 

The design of our new measure relies on the fact that ram pressure has a direct impact on the morphology of the gas component (as traced out by a recently formed stellar population). As a result, the younger stellar population tends to change shape and morphology with respect to that of an intermediate age population of stars.

We present the Size-Shape Difference (SSD) measure that is very sensitive to differences in the size and shape of the younger stellar population when compared to the size and shape of an intermediate age population. Across the parameter space we consider in this study, we find:

   \begin{enumerate}
      \item The SSD measure can easily distinguish between a disk undergoing our fiducial ram pressure and the various tidal interactions we consider. This remains the case for a wide range of ram pressures, as long as the gas disk is truncated inside of the stellar disk. 
      \item The SSD measure can detect the presence of ram pressure even more sensitively when it arises in combination with a tidal interactions. This property is beneficial as it is expected that tidal interactions and ram pressure could occur simultaneously in dense environments.
      \item The best method to detect ram pressure cases is found when combining the SSD measure with a measure of the spiral arm strength in the galaxy disk (in our case measured by the number of detected spiral arms). In this 2D parameter space, we are able to more cleanly separate our tidal interaction, ram pressure and control galaxies over a broad range of times in the simulation, and can even cleanly detect weak ram pressure when combined with a tidal interaction.
   \end{enumerate}

So far we have only tested out the SSD measure in the realm of simulations, in the hope of later developing a measure that could be applied more directly to observations. Therefore it is interesting to ask the question, could our existing recipe be directly applied? In principle, IFU data can and has been used to select stellar populations in specific age bins such as those chosen here for our young and intermediate age populations. However, given the uncertainties, clearly this selection cannot be expected to be perfect. Fortunately, the effectiveness of the current SSD measure was found to not be highly sensitive to the exact choice of age bin for the younger or intermediate age bin. Thus, we believe that the SSD measure could indeed be of value in detecting galaxies suffering ram pressure in IFU observations, and we are currently testing this possibility on real IFU data. Furthermore, our preliminary tests with a $<50$~Myr young population versus a $<300$~Myr intermediate population provide the first indications that we might be able to build an SSD measure for broadband filters (e.g., u or g band versus r or i band). Nevertheless we defer a detailed study of how to apply the SSD more directly to observations to a future study (Lassen et al. (submitted)), where we will be able to more accurately mock the observations, and apply our developed recipes to real IFU and broad band observations of galaxies, and generalize it for galaxies of varying sizes and inclinations.

\begin{acknowledgements}
RS acknowledges financial support from FONDECYT Regular projects 1230441 and 1241426, and also gratefully acknowledges financial support from ANID -- MILENIO -- NCN2024\_112. BV acknowledges financial support from the INAF 2023 GO Grant "Identifying ram pressure induced unwinding arms in cluster spirals" (PI. Benedetta Vulcani). BP acknowledges financial support from the European Research Council (ERC) under the European Unions Horizon 2020 research and innovation program (grant agreement No. 833824, GASP project). YLJ acknowledges support from the Agencia Nacional de Investigaci\'on y Desarrollo (ANID) through Basal project FB210003, FONDECYT Regular projects 1241426 and 1230441, and ANID -- MILENIO -- NCN2024\_112. 
\end{acknowledgements}

%
%

\bibliography{bibfile}{}
\bibliographystyle{aa}

%

\appendix

\onecolumn

\section{Varying the Wind Angle}
\label{sec:app_winddir}

In the main paper, we consider only cases of edge-on ram pressure stripping. In this appendix, we additionally consider cases where the ram pressure wind is inclined to the disk plane by 30$^\circ$ and 90$^\circ$ (`RP-30deg' and `RP-faceon' respectively).

We find that these more inclined winds tend to result in a more heavily truncated gas disk. This is because, in the case of an edge-on wind direction, the leading edge of the disk partially shields the rest of the disk. As a result of the more heavy truncation of the gas disk, the SSD measure responds strongly to the change in the size of the star-forming disk, reaching values significantly in excess of the edge-on case (`RP-fid'). In the case of a face-on wind, which flows along our line-of-sight, this is primarily due to the change in size of the star forming disk as opposed to ram pressure induced asymmetries.

However, following this large peak, the SSD value drops off with time, eventually approaching the values measured for the isolated control model (gray line). This occurs because the stronger ram pressure of the inclined winds results in a small, heavily truncated and round gas disk that is rapidly formed. Up until 200 Myr after the formation of the small star-forming disk, it is only seen in the young age populations. But subsequently, stars formed within the truncated disk will begin appearing in the intermediate age bin (star ages from 200-400 Myr). If we wait until 400 Myr after the disk was truncated, the intermediate age disk will then have a similar size as the young stellar populations. As a result, the SSD drops to low values, due to the similarity in the size and shape of the young and intermediate age populations at these later times.

This appendix demonstrates two key points. Firstly, for a strongly truncated gas disk, in order to detect cases of ram pressure stripping that occurred more than several hundred Myr ago, we should choose an older age range for the intermediate age population (e.g., 400-600 Myr or 600-1000 Myr, instead of 200-400 Myr). Secondly, the SSD remains effective in detecting the presence of ram pressure, even when the ram pressure wind is inclined with respect to the plane of the disk.

\begin{figure}[H]
    \centering
    \includegraphics[width=90mm]{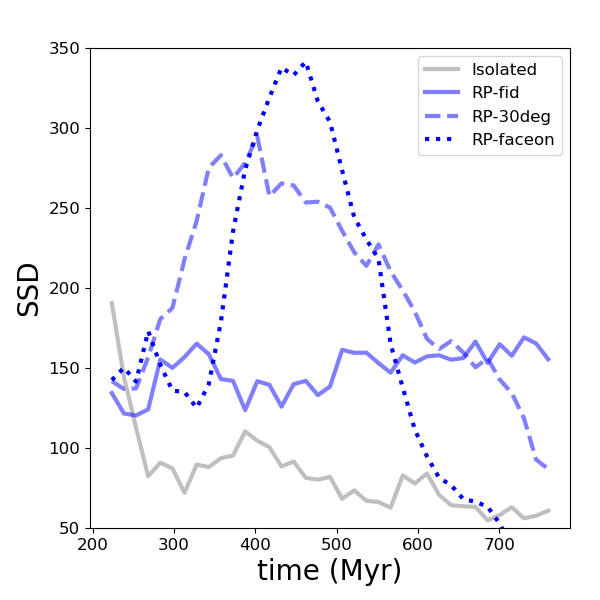}
    \caption{SSD evolution of a single model galaxy for differing choices of the inclination of the ram pressure wind to the plane of the disk. The fiducial model suffers an edge-on wind (`RP-fid'). We also consider a 30$^\circ$ inclined wind (`RP-30deg'), and a wind that is face-on (90$^\circ$ inclined) to the disk plane (`RP-faceon'). We also compare to the isolated control model evolved over the same period (`Isolated'). }
    \label{fig:varywinddirection}
\end{figure}

\clearpage

\section{A Time Evolving Ram Pressure Model}
\label{sec:app_RPSevol}
In the main draft, we limited our modeling of ram pressure to the case of a constant strength ram pressure wind. In this appendix, we consider an example of a time evolving ram pressure wind.

The evolution of the ram pressure is chosen to mimic the infall of a galaxy into a cluster that is similar to the Virgo cluster. For the intracluster medium density profile, we chose a spherical beta-model with $\beta=0.5$, r$_\textrm{core}$=50~kpc, and $\rho_0 = 2 \times 10^{26}$~g/cc. Assuming hydrostatic equilibrium with an isothermal intracluster medium with temperature of $4.7 \times 10^7$ K defines the gravitational potential of the cluster. We then trace the orbit of the infaller in the cluster potential from an initial radius of 1500 kpc, with an initial radial and tangential velocity of 250 km/s. This would result in a typical plunging orbit with a pericentre distance of 185 kpc after 1.6~Gyr. Although, in practice we halt the simulation after 1.25~Gyr of evolution. The resulting ram pressure evolves from a value that is initially 10\% of the RP-fid value. It reaches double that of the RP-fid model by the end of the simulation.

Our set-up is identical to the RP-fid model, only now with a time-evolving ram pressure. In Fig. \ref{fig:RPS_timeevol}, we see the resulting evolution of the SSD measure (`RP-evol'). The initially weak ram pressures result in SSD values that are only marginally above those of the isolated model (gray line). However, as the ram pressure strengthens, from about t$_\textrm{sim}$=600~Myr onwards, the SSD values clearly separate from the isolated model. At t$_\textrm{sim}$=970~Myr, the ram pressure reaches the strength of the RP-fid model. At this moment, the SSD value is about 125 which is less than the values seen in the RP-fid model (typically about 150, see Fig. \ref{fig:varywinddirection}). This SSD value is likely not as large due to the fact that a smoothly changing ram pressure is less shocking to the disk than when a constant ram pressure wind suddenly collides with the disk for the first time. If the gas disk is not as heavily truncated, then it is not surprising that the SSD value is not so strongly perturbed. Nevertheless, the SSD values are still considerably elevated above that of the isolated model (in contrast to our tidal models, see second panel of Fig. \ref{fig:varystuff}). Furthermore, once the ram pressure surpasses the RP-fid level, the SSD value reaches and even surpasses the typical values seen in the RP-fid model.

\begin{figure}[H]
    \centering
    \includegraphics[width=90mm]{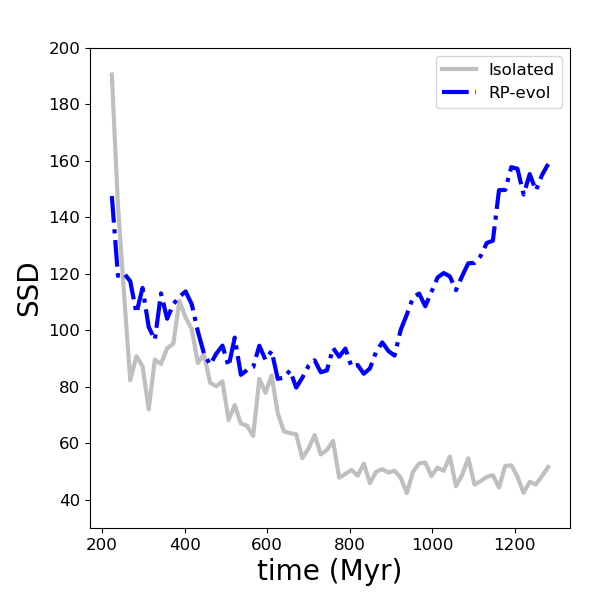}
    \caption{Evolution of the SSD in the case of a time-evolving ram pressure (blue curve) as described in the appendix text. The gray curve is the isolated model for comparison. }
    \label{fig:RPS_timeevol}
\end{figure}

\newpage

\section{Time sequence of TI-fid (left) versus RP-fid (right)}
\label{sec:fid_timeseq}

\begin{figure}[b!] 
    \centering
    \includegraphics[width=80mm]{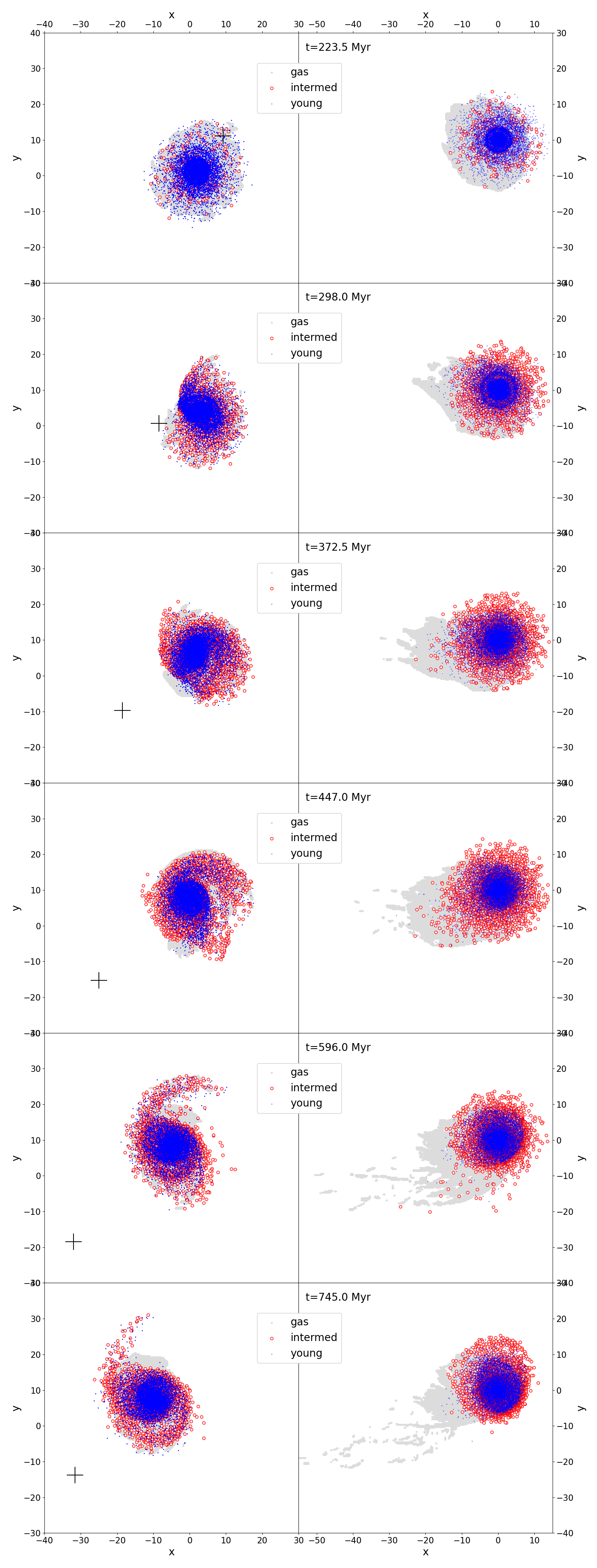}
    \caption{Evolution of the x-y distribution of the young stars (blue points, age$<$200~Myr) and the intermediate age stars (red circles, age 200--400~Myr) for the TI-fid model (left column) and the RP-fid model (right column). From top to bottom, the simulation time increase as indicated in the top-left corner of the panels on the right. The x-y distribution of the disk gas (density$>0.01$~H/cc) is shown with gray shading. In the left column, a cross symbol marks the central position of the secondary galaxy that causes the tidal perturbation in the TI-fid model. In the right column, the ram pressure wind blows from right to left, along the x-axis.}
    \label{fig:fid_timeseq}
\end{figure}

\clearpage

\section{Neglecting the outer Lagrangian radii}
\label{sec:no95}

In this appendix, we present the evolution of the SSD value in the RP-fid model but we exclude the two outermost Lagrangian radii (the 95\% and 99\% Lagrangian radius) from the measure. This is to test how significant the outer regions of the galaxy are for detecting the presence of ram pressure.

In the upper panel, we show the time evolution of the SSD value when all the Lagrangian radii are used for the isolated model (labeled `Isolated') and RP-fid model (`RP-fid'). We compare this with the SSD values when the 95\% and 99\% Lagrangian radii are excluded (`Isolated\_no95' and `RP-fid\_no95'). 

We note that we increase the SSD values by a factor of $5/3$ when excluding the outermost Lagrangian radii, to account for the fact that we now only consider 3 out of the 5 original Lagrangian radii when we calculate the value of the SSD. 

It is clear that, despite this correction, the SSD value of the RP-fid model has been reduced when we exclude the outermost Lagrangian radii. But, importantly, so has the SSD value of the isolated model. Thus, although the SSD of `RP-fid\_no95' is lower than that of `RP-fid', it is still always above the `Isolated\_no95' curve. This means that we are still able to detect the presence of ram pressure without the outermost Lagrangian radii, although the signal appears weaker. This is because, even if only the outer disk is truncated in size, the inner Lagrangian radii can be reduced. This occurs because the Lagrangian radii are percentages of the total light, and the total light has been reduced by the truncation.

To better quantify the weakening of the signal, in the lower panel we plot the ratio of their SSD values  (ram pressure stripped model's SSD value divided by the isolated model's SSD value). When all the Lagrangian radii are used (solid line), the SSD value is typically 1.94 times larger in the case of ram pressure. When the outermost radii are excluded it is 1.47 times larger. The mean decrease in the ratio by excluding the outermost Lagrangian radii is 22\%. We also note that the early peak in SSD value that is seen when the RP-fid model shows the most cometary-like appearance (from time=275--380~Myr) has been removed when we exclude the outermost Lagrangian radii.

These results demonstrate that the outermost Lagrangian radii are important for the SSD measure. They strengthen the signal produced by ram pressure, and may be of particular importance at the early stages of ram pressure. Nevertheless, the SSD measure continues to detect the presence of ram pressure, even in the absence of the outermost Lagrangian radii.

\begin{figure*}[!b] 
    \centering
    \includegraphics[width=100mm]{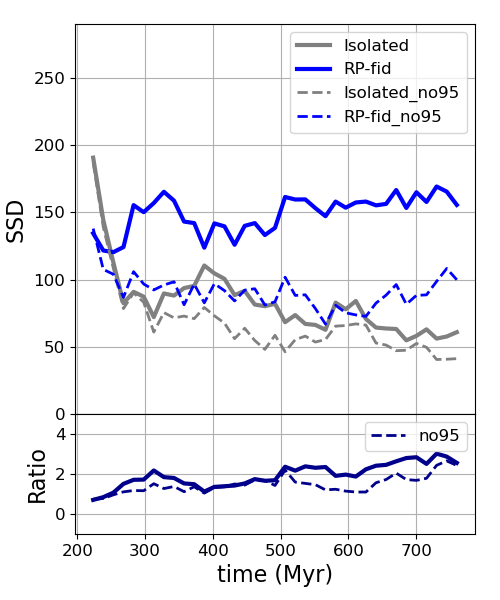}
    \caption{Top panel: Evolution of the SSD value for the isolated (gray curves) and RP-fid (blue curves) model. Solid lines shows the SSD value using all the Lagrangian radii. Dashed lines shows the SSD values excluding the two outermost Lagrangian radii (additionally labeled `\_no95' in the legend). Bottom panel: Ratio of the SSD value of RP-fid model divided by the SSD value of the isolated model. The solid curve is with all Lagrangian radii, the dashed is excluding the two outermost Lagrangian radii (labeled `no95' in the legend).}
    \label{fig:no95}
\end{figure*}

\clearpage

\section{More Extreme Tidal Interaction Models}
\label{sec:app_extrememodels}
In this appendix, we describe two additional tidal interaction models; TI-massive (where the secondary galaxy is more massive, with a 1:2 mass ratio), and TI-fast (where the secondary galaxy moves at 800 km/s, resulting in a high speed tidal encounter. These two tidal interaction models were neglected from the main text to simplify the figures in the main draft, and because their inclusion did not change our main conclusions. Here we show the key results from these models. The large mass of the secondary galaxy in the TI-massive model results in a strong S-shaped structure in the disk of the primary galaxy (see left panel). However, the strong S-shape is shared by both the young (blue points) and intermediate age (red circles) stars, meaning the SSD values only increases slightly compared to the isolated model, with a mean values of SSD=104. Furthermore, the strong S-shape is clearly visible with a typical mean N$_\textrm{peak}$ value of two, further distinguishing it from a case of ram pressure. In the high speed tidal interaction (TI-fast, right panel), the short duration of the tidal interaction causes a negligible disturbance to the disk of the primary galaxy. As a result, in the center panel, the TI-fast points share a similar distribution to the isolated model in the mean N$_\textrm{peak}$ versus SSD diagram.

\begin{figure}[H]
    \centering
    \includegraphics[width=180mm]{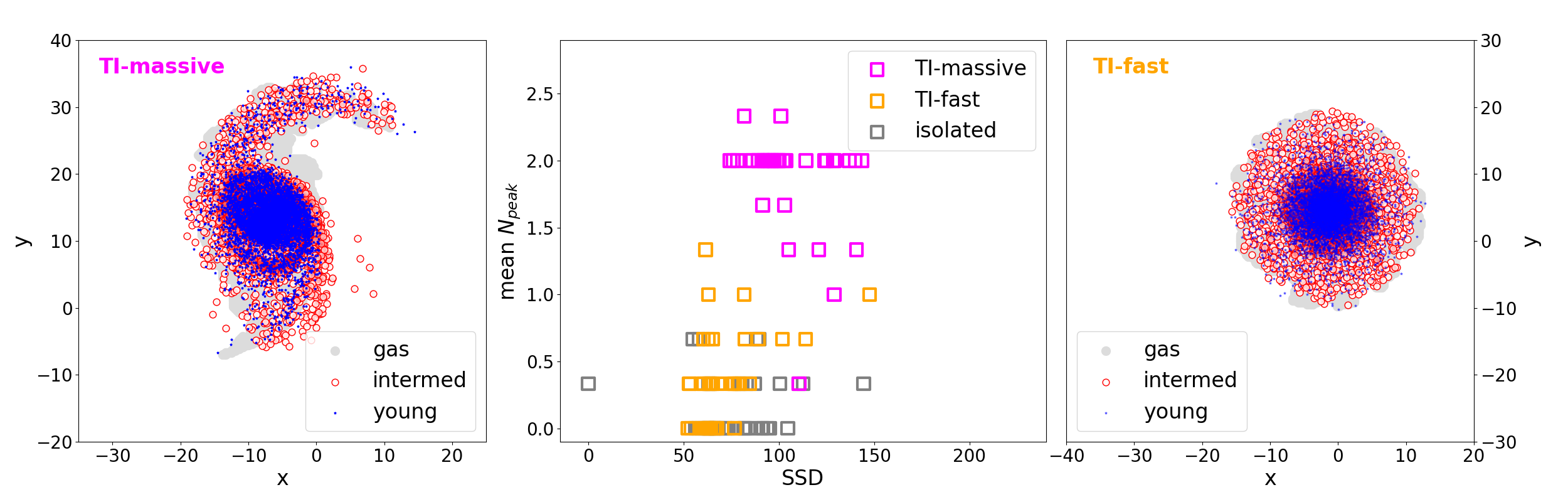}
    \caption{Evolution of the x-y distribution of the young stars (blue points, age$<$200~Myr) and the intermediate age stars (red circles, age 200--400~Myr) for the TI-massive model (left panel) and the TI-fast model (right column), shown at t$_\textrm{sim}$=522 Myr. The center panel shows the values of mean N$_\textrm{peak}$ (y-axis) versus the SSD for each snapshot with t$_\textrm{sim}>$220 Myr of the TI-massive (purple) and TI-fast (orange) model. The isolated model (gray) is shown for comparison over the same time period.}
    \label{fig:2Dextremes}
\end{figure}

\clearpage

\section{The CAS parameters}
\label{sec:CASstuff}

In this appendix, we calculate the Concentration, Asymmetry and Smoothness parameters (the so called `CAS' parameters; \cite{Conselice2001}) for the TI-fid and RP-fid model, and compare them with the isolated model. The CAS values are measured on the young stars (age $<200$ Myr). 

Concentration (C) is defined as R80/R20 (the radius containing 80$\%$ of the stars divided by the radius containing 20$\%$ of the stars). As ram pressure truncates the gas disk, the young stars shift to smaller radii. This causes a small increase in the C-parameter above that of the the isolated model. Meanwhile the TI-fid model gives very similar C-values to the isolated model.

The Asymmetry (A) is measured by flipping the image of the model galaxy's young stars 180 degrees along the x-axis, and subtracting the flipped image from the original. The absolute value of the pixel-by-pixel differences are summed across all pixels of the image and divided by the total flux in the image to give the final A-value. Due to low-N statistics of the young stars, we conduct a 1 kpc wide tophat smooth of the image to suppress graininess before measuring the A-value. As the cometary shape of the RP-fid model is not symmetric, and the spiral arms of the TI-fid model are not symmetric either (e.g., see Fig. \ref{fig:agediff}), the A-value of both RP-fid and TI-fid is increased with respect to the isolated model. However, the increase is generally quite small ($\sim$0.1) and both TI-fid and RP-fid share similar values, which makes it difficult to distinguish between them with the A-value. 

The Smoothness value (S) is measured by smoothing the image (with a 2 kpc wide tophat smooth) and then subtracting the smoothed image from the original image. The remaining residual is then normalized by the total flux in the image, and subtracted from unity (in order to have higher smooth values for smoother disks). In practice, we conduct a single 1 kpc wide tophat pre-smooth of the image, in order to remove graininess due to low numbers of star particles (as we did with the Asymmetry value) before measuring the smoothness. As ram pressure does not generate non-smooth structure in the disk stars of the RP-fid model, the S-values are very similar to the isolated model. However, the S-shaped structure induced in the TI-fidk model's disk causes a small increase in the S value above those of the isolated model.

In summary, the response of the Concentration and Smoothness parameter is quite weak to the ram pressure and tidal interactions that occur in the RP-fid and TI-fid model, respectively. The response of the Asymmetry parameter is slightly more significant. However, both RP-fid and TI-fid respond in a similar way. This means the SSD is of more value than the Asymmetry parameter for distinguishing the influence of ram pressure from the influence of tidal interactions.

\begin{figure*}[!b] 
    \centering
    \includegraphics[width=120mm]{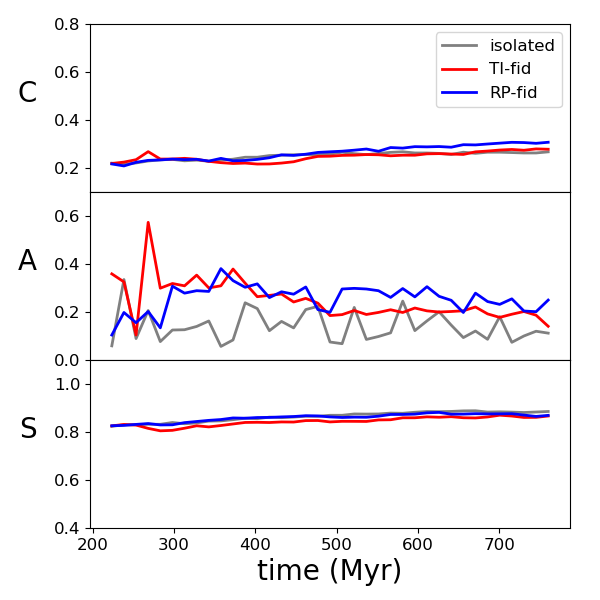}
    \caption{Time evolution of the Concentration parameter C (top panel), the Asymmetry parameter A (middle panel), and the Smoothness parameter S (lower panel), for the TI-fid, RP-fid and isolated model (see legend)}
    \label{fig:CAS}
\end{figure*}

\clearpage


\end{document}